\documentclass[aps,prb, preprint]{revtex4}
\usepackage{amsmath,graphicx}

\def\bra#1{\left\langle #1 \right|}
\def\ket#1{\left| #1 \right\rangle}

\begin{document}

\title{Exciton luminescence in resonant photonic crystals}
\author{L. I. Deych$^{\dag}$}
\author{M. V. Erementchouk$^{\ddag}$}
\author{A. A. Lisyansky$^{\dag}$}
\author{E. L. Ivchenko$^{\S}$}
\author{M. M. Voronov$^{\S}$}

\affiliation{$^{\dag}$ Physics Department, Queens College, City
University of New York, Flushing, New York 11367, USA}

\affiliation{$^{\ddag}$Department of Physics and Astronomy,
Northwestern University, Evanston, IL 60208}

\affiliation{$^{\S}$
A.F. Ioffe Physico-Technical Institute, Politekhnicheskaya str.~26,
St. Petersburg, 194021, Russia}

\begin{abstract}
A phenomenological theory of luminescence properties of
one-dimensional resonant photonic crystals is developed within the
framework of classical Maxwell equations with fluctuating
polarization terms representing non-coherent sources of emission.
The theory is based on an effective general approach to determining
linear response of these structures and takes into account formation
of polariton modes due to coherent radiative coupling between their
constituting elements.  The general results are applied to Bragg
multiple-quantum-well structures, and theoretical luminescence
spectra of these systems are compared with experimental results. It
is shown that the emission of such systems can be significantly
influenced by deliberately introducing defect elements in the
structure. The relation between absorption and luminescence spectra
is also discussed.
\end{abstract}

\maketitle

\section{Introduction}

A possibility to influence the emission of light by tailoring the
dielectric environment of emitting objects has been attracting a
great deal of attention since a pioneering work by Yablonovitch.
\cite{YABLONOVITCH:1987} In Ref.~\onlinecite{YABLONOVITCH:1987} it
was suggested that a three-dimensional periodic modulation of the
dielectric constant can result in formation of a photonic band
structure, consisting of allowed and forbidden photonic bands in
analogy with electronic band structure. One of the important
consequences of the band structure is a modification of
electromagnetic density of states, which can be used to suppress or
enhance the rate of the spontaneous emission of emitters embedded in
a photonic crystal. While such drastic effects as the full
inhibition of the spontaneous emission proved to be difficult to
achieve,\cite{Joannopoulos_PC_MFL:1995,WOLDEYOHANNES:2003} there is
still a growing interest in emission properties of photonic
crystals,\cite{PhysRevA.59.4727,{PhysRevB.63.125106},{PhysRevA.64.033801},{PhysRevA.65.043808},{barth:243902}}
which even in the absence of complete photonic band-gap may result
in a significant modification of properties of emitted radiation. If
one is not looking to achieve the full inhibition of the spontaneous
emission, the systems in which a periodic modulation takes place in
only two or even one dimension are also of interest, because even
though they cannot confine light completely, they do modify emission
patterns for particular directions, which can be useful for various
applications.

One-dimensional structures attract a particularly great attention,
firstly, because they are easiest to manufacture, and secondly,
because they allow for a detailed theoretical description. These two
circumstances make one-dimensional structures most suitable
candidates for a number of applications that do not require
modification of the electromagnetic properties in all three
dimensions. At the same time, certain properties of one-dimensional
structures are typical for two- and three-dimensional systems as
well, and, therefore,  these structures provide a convenient testing
area for understanding some general properties of media with spatial
modulation of the dielectric function.
%One-dimensional structures are easiest to create from technological
%point of view, and also simplest for theoretical analysis.
%Therefore, they have been attracting a particularly great attention.

Most of the previous works addressed the problem of the spontaneous
emission in 1D photonic structures from the perspective of
individual emitters embedded in special dielectric environments such
as  superlattices or a Fabry-Perot cavities (see, for instance,
Refs.~\onlinecite{BRUECK:2003,HOEKSTRA:2005,Savona:1996,Hayes:1998}
and references therein). Theoretical analysis of these situations is
based on the assumption that the structure of photonic modes is
determined solely by the periodical modulation of the dielectric
function. The interaction between  the photonic modes and emitters
is considered in this approach as weak in a sense, that it does not
affect the structure of the photonic modes, and can be treated
perturbatively within the framework of Fermi Golden Rule.

These assumptions, however, break down in the important case of so
called resonant photonic crystals, that have begun attracting
considerable attention in recent years. These structures are
composed of periodically distributed structural elements containing
dipole active internal excitations, which co-exist with periodic
modulation of
the dielectric constant.\cite{KUZMIAK:1997,DEYCH:1998,NOJIMA:2000,%
Raikh_braggaritons,CHRIST:2003,HUANG:2003,TOADER:2004,PILOZZI:2004,IVCHENKO:2005,JointComplex}
Multiple-quantum-wells\cite{IvchenkoMQW} (MQW) present one of the
popular realizations of such structures with the one-dimensional
periodicity. Excitons confined within quantum wells provide
optically active excitations, and the contrast between refractive
indices of wells and barriers is responsible for periodic modulation
of the dielectric constant. When the period of the structure
satisfies a special, so called, Bragg condition, the interaction
between light and excitons cannot be considered as weak, and cannot
be treated with the help of the Fermi golden rule. Excitons and
periodic modulation of the refractive index play in such structures
equally important roles in the formation of the photon modes, which
should be more appropriately called polaritons. The Bragg condition,
in its most general form can be written as
$\Delta\phi_p(\omega_{0})=\pi$, where $\Delta\phi_p$ is the change
of the phase of the propagating electromagnetic wave over one period
of the structure calculated at the exciton frequency
$\omega_0$\cite{DispersionMQWPC}. If one neglects the refractive
index contrast (optical lattice approximation) this condition can be
rewritten as $c\omega_0/d=\pi$,\cite{IvchenkoMQW} where $d$ is the
period of the structure, and $c$ is the speed of light in the
medium.

%By now dispersion properties and reflection spectra of the Bragg
%structures have been intensively studied both theoretically
%\cite{IvchenkoSpectrum,JointComplex,PILOZZI:2004,Optical_Properties_MQWPC,DispersionMQWPC}
%and experimentally. \cite{HUBNER:1999,time_resolved_contrast,Hayes}

The emission of light in such structures differs significantly from
the cases considered in
Refs.~\onlinecite{Savona:1996,Hayes:1998,BRUECK:2003,HOEKSTRA:2005}.
The emitters of light in resonant photonic crystals affect the
spatial structure of electromagnetic modes as much as the modulation
of the refractive index. As a result, the processes of light
emission by a particular quantum well and its propagation inside the
structure should be considered on equal footing. To develop a
theoretical formalism for dealing with such situations is the main
objective of this paper. While focusing on the luminescent
properties of Bragg MQW structures, we will treat them in a broad
context of resonant photonic crystals. This allows us to develop a
universal theoretical formalism, applicable to essentially any type
of one dimensional structures with periodically distributed
emitters.

Since we will be interested in the effects of photonic environment
on the luminescence rather than in a microscopical description of
the processes of the exciton relaxation and recombination, we will
base our theory on macroscopical Maxwell equations with a
non-coherent polarization source term, which would simulate
non-coherent exciton population created by a non-resonant pumping.
The phenomenological nature of our work distinguishes it from recent
paper \onlinecite{KIRA:2006}, in which microscopic theory of
spontaneous emission of Bragg MQW structures was based on an
approach, in which both electron-hole dynamics and electromagnetic
field were treated quantum-mechanically\cite{KIRA:1999}.
Phenomenological nature of our theory allows for establishing  a
direct and clear connection between global optical properties of MQW
structures and their luminescence spectra, which is important for
understanding  an intimate relationship between geometrical
structure of MQW systems and their emission properties. On the other
hand, properties of the polarization source entering our
calculations as a phenomenological object can only be established
either from independent experiments or on the basis of microscopical
theories of the kind developed in Ref. \onlinecite{KIRA:2006}.

The theoretical formalism developed in this paper solves also a more
general problem of calculating an optical linear response of finite
1D resonant photonic crystals. Usually, the linear response is
studied using Green's function formalism, based on spectral
representation of the Green's function. The later, however, is not
very well suited to deal with finite structures whose normal modes
cannot be considered as eigenfunctions of a Hermitian operator. In
our approach we develop a method of relating Green's function of the
finite resonant photonic crystal to transfer matrices describing
reflection and transmission properties of these structures.

The paper has the following structure. In
Section~\ref{sec:Macroscopic_Maxwell_equations} we formulate the
basic equations describing the distribution of the electric field in
the structure and introduce the basic elements of our transfer
matrix approach. In Section~\ref{sec:Greens_function} we conclude
the presentation of the general formalism by deriving a general
expression for the respective Green's function. In
Section~\ref{sec:Emission_spectrum} the general formalism is applied
to the problem of the exciton luminescence spectrum in resonant
photonic crystals. In addition to considering an ideal periodic
structure, we also discuss the role of inhomogeneous broadening,
modifications in the luminescence induced by a deliberate
introduction defects in the otherwise periodic structure, and
relation between luminescence and absorption spectra. The paper is
concluded with Appendix, where we discuss relation between our
phenomenological approach and  microscopic theories.

\section{General solution of Maxwell equations in one-dimensional resonant photonic crystal with polarization sources}
\label{sec:Macroscopic_Maxwell_equations}
\subsection{Maxwell equations for multiple-quantum-well system with
sources of polarization}

Our approach to description of emission properties of MQW structures
is based on solution of classical Maxwell equations for
monochromatic field with frequency $\omega$ of the following form:
\begin{equation}\label{eq:Mawell_equation_start}
 \nabla \times \nabla \times \mathbf{E} = \frac{\omega^2}{c^2}\left[ n^2(z)\mathbf{E}
 + 4\pi \mathbf{P} _{\mathrm{exc}}+4\pi\mathbf{F}\right],
\end{equation}
where the coordinate $z$ is chosen to represent the growth direction
of the structure, and $n(z)$ is the periodically modulated
background index of refraction: $n(z+d) = n(z)$. The polarization in
this equation is presented as a sum of two terms. First of them,
$\mathbf{P}_{\mathrm{exc}}$, originates from optical transitions,
with frequencies within the spectral region of interest and has,
therefore resonant behavior. The second contribution,
$\mathbf{F}({\bf r})$, arises  due to other emitting transitions,
which are non-resonant at given frequencies. In typical experimental
situation involving $III-V$ quantum wells, the resonant term
corresponds to polarization due to $1s$ heavy-hole excitons, and the
non-resonant contribution can be considered as a contribution from
all other optically active transitions between states of
electron-hole system. Such a separation of polarization into
contributions from different optical transitions is possible if one
neglects Coulomb correlations between exciton states and
electron-hole plasma. These correlations are important for for
highly excited states of semiconductors\cite{KIRA:1999}, but can be
neglected if concentration of photo-excited electron-hole pairs is
not too high, which is the situation considered in this work.

Semi-classical description of light-exciton interaction in a single
quantum well (see, for instance, Ref.\onlinecite{Ivchenko:Book})
indicates that the exciton contribution to polarization of a single
QW can be presented in the following form:
\begin{equation}\label{eq:exciton_polarization_singleQW}
 P_{\mathrm{exc}}^{(m)} = -\chi_m(\omega)\Phi_m(z) \left[ \int
 dz'\,\Phi_m(z')\mathbf{E}_\perp(z',\boldsymbol{\rho}) +
 \boldsymbol{\Sigma}_m(\boldsymbol{\rho})\right].
\end{equation}
Here index $m$ numerates quantum wells, $\boldsymbol{\rho}$ is
two-dimensional position vector perpendicular to the growth
direction of the structure, and the wave function of the exciton
localized in the $m$-th well, $\Phi_m(z)$, is taken in the form
$\Phi_m(z) = \Phi(z-z_m)$, where $z_m$ is the position of the center
of the $m$-th well. The expression proportional to the electric
field in this equation describes direct optical excitation of
excitons, while the second term, represented by function
$\boldsymbol{\Sigma}_m(\boldsymbol{\rho})$, introduces an additional
source of exciton polarization due to non-radiative processes.
Depending on the properties of this term it can describe either
coherent or non-coherent emission. More detailed description of
function $\boldsymbol{\Sigma}_m(\boldsymbol{\rho})$ in relation to
the description of luminescence, as well as general justification of
the phenomenological approach to this problem is given below in
Section \ref{sec:Emission_spectrum}. Here we will treat
$\boldsymbol{\Sigma}_m(\boldsymbol{\rho})$ as an arbitrary function
responsible for generation of exciton polarization.

Writing Eq.~(\ref{eq:exciton_polarization_singleQW}) we explicitly
take into account that the dipole moment of heavy-hole excitons is
oriented in the plane of the well and, therefore, only components of
the field perpendicular to the growth direction, $\mathbf{E}_\perp$,
enter the expression for coherent exciton polarization. The
intensity of the exciton-light interaction is characterized by the
exciton susceptibility $\chi_m(\omega)$. Neglecting the exciton
dispersion in the plane of the quantum well and the inhomogeneous
broadening the susceptibility can be written in the form
\begin{equation}\label{eq:exciton_susceptibility}
  \chi_m(\omega) = \frac{\alpha}{{\omega_m}_0 - \omega - i\gamma}.
\end{equation}
where $\alpha $ is the exciton-light coupling parameter proportional
to the exciton dipole moment, ${\omega_m}_0$ is the exciton
resonance frequency in the $m$-th well, and $\gamma$ is the
homogeneous broadening of the exciton line.

The exciton polarization of the entire MQW system is the sum of
polarizations of individual wells:
\begin{equation}\label{eq:exciton_polarization_start}
 P_{\mathrm{exc}} = \sum_m P^{(m)}.
\end{equation}
Here we assume that the period of the spatial arrangement of the
quantum wells coincides with the period of the modulation of the
dielectric function, such that the distance between adjacent wells
is $z_{m+1}- z_m =d$. In principle, the non-resonant polarization
term is also a sum of contribution from different wells, but since
it does not depend on electric field, it is more convenient to keep
it in the equation as a single term.

Maxwell equation (\ref{eq:Mawell_equation_start}) together with
polarization given by Eq.~(\ref{eq:exciton_polarization_singleQW})
and (\ref{eq:exciton_polarization_start}) describes a system of
quantum well excitons interacting with common radiative field, ${\bf
E}$. In the absence of polarization sources these equations have
been intensively studied, and it is well known that they describe
collective dynamics of radiatively coupled  QW
excitons\cite{IvchenkoMQW,{IvchenkoSpectrum},{HUBNER:1999}}. The
main objective of the present section is to develop a general
theoretical approach to solving
Eq.~(\ref{eq:Mawell_equation_start}),
Eq.~(\ref{eq:exciton_polarization_singleQW}) and
(\ref{eq:exciton_polarization_start}) in the presence of the sources
of polarization of an arbitrary form. However, since in this paper
the developed approach will be mainly applied to exciton
luminescence spectrum in the growth direction of the structure, we
will restrict, for simplicity, our consideration only to
$s$-polarized radiation.

Using the translational invariance of the system in the $x-y$ plane,
we can present solutions of  Eq.(\ref{eq:Mawell_equation_start}) in
the form
\begin{equation}\label{eq:field_separation_variables}
  \mathbf{E}(z,\boldsymbol{\rho}) = e^{i\mathbf{k}\boldsymbol{\rho}}
  \mathbf{E}(z,\mathbf{k}),
\end{equation}
where  $\mathbf{k}$ is the in-plane wave vector. For a $s$-polarized
wave the direction of $\mathbf{k}$ determines the direction of
$\mathbf{E}(z,\mathbf{k})$ as
\begin{equation}\label{eq:field_scalar_amplitude}
  \mathbf{E}(z,\mathbf{k}) = E(z,\mathbf{k})
  \hat{\mathbf{e}}_s(\mathbf{k}),\qquad \hat{\mathbf{e}}_s(\mathbf{k})
\equiv\hat{\mathbf{e}}_z\times\hat{\mathbf{e}}_\mathbf{k}
\end{equation}
where $\hat{\mathbf{e}}_s$, $\hat{\mathbf{e}}_z$, and
$\hat{\mathbf{e}}_\mathbf{k}$ are unit vectors describing directions
of polarization, growth direction and the direction of the in-plane
wave vector, respectively. In what follows we will omit the argument
$\mathbf{k}$ when it is clear from the context that the value of the
scalar amplitude is taken at a fixed value of the in-plane wave
vector.

Substituting  Eq.~(\ref{eq:field_separation_variables}) into Maxwell
equation (\ref{eq:Mawell_equation_start}) and choosing $s$-polarized
component of the field according to representation
(\ref{eq:field_scalar_amplitude}), we derive the following equation
for the scalar amplitude of the field
\begin{equation}\label{eq:field_scalar_equation}
\begin{split}
 \frac{d^2 E(z)}{dz^2} + &\kappa^2(z)E(z) = \frac{4\pi\omega^2}{c^2}F(z) -\\
 \frac{4\pi\omega^2}{c^2}&\sum_m\chi_m(\omega)\Phi_m(z)\left[\int dz'\,\Phi_m(z')E(z')+\Sigma_m\right],
\end{split}
\end{equation}
where $\kappa^2(z) = \omega^2 n^2(z)/c^2 - k^2$, and $\Sigma_m$,
$F(z)$ are the components of the two-dimensional Fourier transforms
of the source terms $\boldsymbol{\Sigma}_m(\boldsymbol{\rho})$ and
$\mathbf{F}({\bf r})$ in the direction of $\hat{\mathbf{e}}_s$:
%\begin{equation}\label{eq:Fourier_representation_sources}
%\begin{split}
%  \boldsymbol{\Sigma}_m(\boldsymbol{\rho}) =
%    \int d^2 k\, \hat{\mathbf{e}}_s(\mathbf{k})\Sigma_m(\mathbf{k})
%    e^{i \mathbf{k}\boldsymbol{\rho}}, \\
%%
% F(\boldsymbol{\rho},z) = \int d^2 k\, \hat{\mathbf{e}}_s(\mathbf{k})F(\mathbf{k},z)
%    e^{i \mathbf{k}\boldsymbol{\rho}},
%\end{split}
%\end{equation}
%where the integrals are taken over vectors lying in the plane of
%quantum wells and the coefficients $\Sigma_m(\mathbf{k})$ and
%$F(\mathbf{k},z)$ can be found using the Fourier transformation
\begin{equation}\label{eq:sources_Fourier}
\begin{split}
  \Sigma_m(\mathbf{k}) = \hat{\mathbf{e}}_s(\mathbf{k}) \cdot
    \int d^2 \rho\, \boldsymbol{\Sigma}_m(\boldsymbol{\rho})
    e^{-i \mathbf{k}\boldsymbol{\rho}}, \\
 F(\mathbf{k},z) = \hat{\mathbf{e}}_s(\mathbf{k}) \cdot
    \int d^2 \rho \mathbf{F}(\boldsymbol{\rho},z)
    e^{-i \mathbf{k}\boldsymbol{\rho}}.
\end{split}
\end{equation}
%In what follows we will drop  $\mathbf{k}$ as the argument in
%expressions for $\Sigma_m(\mathbf{k})$ and $F(\mathbf{k},z)$.

Equation (\ref{eq:field_scalar_equation}) is the starting equation
for the formalism developed below. The essential assumptions for
this formalism are the non-local character of the exciton-light
interaction and the possibility to separate in-plane coordinates.
These assumptions are not too restrictive and, therefore, the
formalism can be generalized for more complicated situations such as
multi-resonance form of the susceptibility or the presence of the
inhomogeneous broadening of excitons. The latter can be accounted
for in the effective medium approximation by replacing the exciton
susceptibility (\ref{eq:exciton_susceptibility}) with its averaged
over exciton frequencies version.\cite{OmegaDefectPRB}
%
%It should be noted that the form of the exciton susceptibility does
%not affect the possibility to separate the in-plane coordinates and
%reduce the problem to the effectively one-dimensional situation as
%long as the in-plane translation invariance is preserved. Thus, the
%formalism developed below can be applied to more complicated
%situations such as multi-resonance form of the susceptibility or the
%presence of the spatial dispersion of excitons.
%%We can even take into account inhomogeneous broadening of excitons
%%in the effective medium approximation by replacing the exciton
%%susceptibility (\ref{eq:exciton_susceptibility}) with its averaged
%%over exciton frequencies version\cite{OmegaDefectPRB}.
%The choice of the simplest one-level form of the susceptibility is
%motivated by our intention to present our method of dealing with
%radiative properties of MQW structures without unnecessary technical
%complications.

\subsection{Transfer matrix approach to one-dimensional equations
with sources} \label{sec:Transfer_matrix_formulation}

The reduction of the initial problem to the one-dimensional equation
(\ref{eq:field_scalar_equation}) allows us to solve it using a
powerful transfer-matrix technique. A convenient formulation of this
approach specifically adapted for the structures under consideration
was developed in Ref.~\onlinecite{DispersionMQWPC}. The presence of
the source terms in Eq.~(\ref{eq:field_scalar_equation}), however,
requires some modifications of that approach, and the adaptation of
the transfer-matrix method to inhomogeneous integro-differential
equations is one of the important technical results of this paper.

Without any loss of generality we can consider a layer with the
quantum well situated at $z=0$ with the left and right boundaries at
$z_-$ and $z_+$ respectively. Inside a single layer the summation
over quantum wells in Eq.~(\ref{eq:field_scalar_equation}) as well
as the well's index, can be dropped, and we can rewrite this
equation in the  form of a second order inhomogeneous differential
equation, in which polarization terms appear as the right hand side
inhomogeneity:
\begin{equation}\label{eq:inhomogeneous_equation}
 \frac{d^2E(z)}{dz^2}+ \kappa^2(z)E(z) = \mathcal{F}(z).
\end{equation}
A general solution of such an equation has the
form\cite{Morse_Feshbach}
\begin{equation}\label{eq:general_solution}
  E(z) =  c_1 h_1(z) + c_2 h_2(z) + (G \star \mathcal{F})(z) ,
\end{equation}
where $h_{1,2}(z)$ are a pair of linearly independent solutions of
the homogeneous equation
\begin{equation}\label{eq:pure_PC_equation}
  \frac{d^2E(z)}{dz^2}+ \kappa^2(z)E(z) = 0,
\end{equation}
and
\begin{equation}\label{eq:convolution}
  (G \star \mathcal{F})(z) = \int_{z_-}^zdz'\,G(z,z')
\mathcal{F}(z')
\end{equation}
Here $G(z,z')$ describes the linear response of a passive (without
exciton resonances) 1D photonic crystal and can be expressed in
terms of functions $h_{1,2}$ as
\begin{equation}\label{eq:PC_Green_function}
G(z,z')=\frac{1}{W_h}\left[h_1(z')h_2(z)-h_1(z)h_2(z')\right],
\end{equation}
where Wronskian, $W_h = h_1 dh_2/dz - h_2 dh_1/dz$,  does not depend
on $z$.

We  choose $h_{1,2}$ as real valued solutions of the Cauchy problem
for Eq.~(\ref{eq:pure_PC_equation}) and use them to present the
electric field at the left boundary of the elementary cell, $z=z_-$,
as
\begin{equation}\label{eq:E_left_boundary}
E(z_-) =  c_1 h_1(z_-) + c_2 h_2(z_-).
\end{equation}

Combining Eq.~(\ref{eq:field_scalar_equation}) with
Eqs.~(\ref{eq:general_solution}), (\ref{eq:convolution}), and
(\ref{eq:PC_Green_function}) we can derive the following expression
for the value of the field at the right boundary of the elementary
cell, $z_+$:
\begin{equation}\label{eq:general_solution_right_boundary}
\begin{split}
 E(z_+) = &h_1(z_+)\left[c_1 -\frac{4\pi\omega^2F_2}{c^2\sqrt{W_h}} + \tilde\chi\frac{4\pi\omega^2\varphi_2}{c^2}
       \left(c_1 \varphi_1 + c_2 \varphi_2 + \frac{\tilde \Sigma}{\sqrt{W_h}} \right)\right]\\
 +& h_2(z_+)\left[c_2 + \frac{4\pi\omega^2F_1}{c^2\sqrt{W_h}} - \tilde\chi\frac{4\pi\omega^2\varphi_1}{c^2} \left(
        c_1 \varphi_1 + c_2 \varphi_2 + \frac{\tilde \Sigma}{\sqrt{W_h}}\right)\right],
\end{split}
\end{equation}
where $\varphi_{1,2}$ and $F_{1,2}$ are the ``projections" of the
exciton state and the non-resonant field source onto the functions
$h_{1,2}$
\begin{equation}\label{eq:projection_of_PC_modes}
\begin{split}
  \varphi_{1,2} = \frac{1}{\sqrt{W_h}}\int_{z_-}^{z_+}dz \Phi(z)h_{1,2}(z), \\
 F_{1,2} = \frac{1}{\sqrt{W_h}}\int_{z_-}^{z_+} dz\, F(z)
 h_{1,2}(z).
\end{split}
\end{equation}
In Eqs.~(\ref{eq:projection_of_PC_modes}) the integrals are taken
over the period of the structure (or over the elementary cell of the
photonic crystal). The effective polarization source function
$\tilde\Sigma$ is the initial $\Sigma$ modified by the field source
function
\begin{equation}\label{eq:modified_Sigma}
  \tilde\Sigma = \Sigma + \int_{z_-}^{z_+} dz\, \Phi(z)(G\star F)(z).
\end{equation}
In Eq.~(\ref{eq:general_solution_right_boundary}) we also have
introduced the modified exciton susceptibility
\begin{equation}\label{eq:modified_susceptibility}
  \tilde\chi = \frac{\chi}{1+\Delta\omega\chi/\alpha},
\end{equation}
where
\begin{equation}\label{eq:radiative_shift}
  \Delta\omega =\frac{4\pi\omega^2\alpha}{c^2}\int_{QW}dz\,
\Phi(z)(G\star\Phi)(z)
\end{equation}
is the radiative correction to exciton susceptibility in the
photonic crystal.

Taking into account that the electric field at $z=z_+$ can also be
presented in the form of Eq.~(\ref{eq:E_left_boundary}) with
modified coefficients $c_{1,2}$, we can describe the evolution of
the field upon propagation across the elementary cell of the
structure as a change in these coefficients. Using the solution
(\ref{eq:general_solution_right_boundary}) the relation between the
coefficients at different boundaries of the elementary cell can be
found as
\begin{equation}\label{eq:amplitudes_transfer}
  \begin{pmatrix}
    c_1 \\
    c_2
  \end{pmatrix}(z_+) = \widehat{T}_h \begin{pmatrix}
    c_1 \\
    c_2
  \end{pmatrix}(z_-) + \begin{pmatrix}
    \Delta c_1 \\
    \Delta c_2
  \end{pmatrix},
\end{equation}
where the two dimensional vectors $(c_1,c_2)(z_+)$ and
$(c_1,c_2)(z_-)$ represent the set of the respective coefficients,
and $\widehat{T}_h$ is the transfer matrix describing their
evolution across the elementary cell written in the basis of the
linearly independent solutions $h_{1,2}$:
\begin{equation}\label{eq:transfer_matrix_functions_basis}
  \widehat{T}_h = \hat{1} + \frac{4\pi\omega^2\tilde\chi}{c^2}
  \begin{pmatrix}
    \varphi_2\varphi_1 & \varphi_2^2 \\
    -\varphi_1^2 & -\varphi_2\varphi_1
  \end{pmatrix},
\end{equation}
$\hat{1}$ is the unit matrix. The contribution of the sources into
the field is described by the second term in r.h.s. of
Eq.~(\ref{eq:amplitudes_transfer})
\begin{equation}\label{eq:amplitudes_luminescence_shift}
\begin{pmatrix}
    \Delta c_1 \\
    \Delta c_2
  \end{pmatrix} = \frac{4\pi\omega^2}{c^2\sqrt{W_h}}\begin{pmatrix}
    -F_2 \\
    F_1
  \end{pmatrix} +
\tilde\Sigma\,\frac{4\pi\omega^2\tilde\chi}{c^2\sqrt{W_h}}\begin{pmatrix}
    \varphi_2 \\
    -\varphi_1
  \end{pmatrix}.
\end{equation}
This is one of the main results of this Section. Its important
feature is that the source contribution is independent of the state
of the ``incoming" field. In other words, the value of the field at
the right boundary of the elementary cell is a superposition of a
field propagated across the elementary cell from the left boundary
(as if there were no sources at all) and the field generated by
sources. This result is of a very general nature and can be applied
to a variety of situations such as multipole exciton susceptibility,
asymmetric quantum wells, incommensurability between periodicity of
quantum well positions and modulation of the refractive index, etc.

It might appear that Eq.~(\ref{eq:amplitudes_transfer}) violates the
symmetry between the left and the right since the sources contribute
only to the field at the right boundary of the elementary cell. This
apparent asymmetry results from a fact that
Eq.~(\ref{eq:amplitudes_transfer}) presents a solution of the Cauchy
problem,  which is inherently asymmetric. The left-right symmetry
should be expected only from a solution satisfying radiative
boundary conditions and in the next section of the paper we
demonstrate how to use Eq.~(\ref{eq:amplitudes_transfer}) to find
such a solution.

\begin{figure}
  \includegraphics[width=3in]{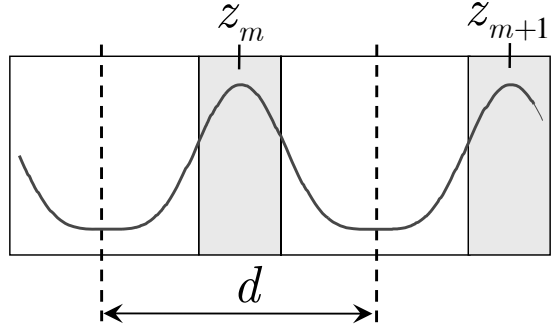}\\
  \caption{The periodic structure built of quantum wells (the shadowed rectangulars)
  and the barriers between them. Vertical
  dashed lines show the boundary of the elementary cell having the property
  of the mirror symmetry. The smooth line illustrates the modulation
  of the dielectric function in the structure.}\label{fig:elementary_cell}
\end{figure}

The expressions presented in Eqs.
(\ref{eq:transfer_matrix_functions_basis}) and
(\ref{eq:amplitudes_luminescence_shift}) can be greatly simplified
in the case of symmetric quantum  wells and the modulation of the
dielectric function, which is consistent with this symmetry, i.e.
$n(z_m+z) = n(z_m-z)$, where $z_m$ is the position of the center of
$m$-th quantum well. In this case the elementary cell of the
structure can be chosen to have the explicit mirror symmetry with
respect to its center (see Fig.~\ref{fig:elementary_cell}), and this
is the case that we consider in what follows.

Because of invariance of Eq.~(\ref{eq:pure_PC_equation}) with
respect to mirror reflection, its solutions can be chosen to have a
definite parity. Thus we can choose linearly independent solutions
$h_{1,2}$ to be either even or odd with respect to the center of the
quantum well. For concreteness we choose $h_2$ to be the odd
solution, which result in $\varphi_2$ turning to zero and transfer
matrix $T_f$ taking the following much simpler form:
\begin{equation}\label{eq:transfer_matrix_functions_basis_simple}
  \widehat{T}_h = \hat{1} + \frac{4\pi\omega^2\tilde\chi}{c^2}
  \begin{pmatrix}
    0 & 0 \\
    -\varphi_1^2 & 0
  \end{pmatrix},
\end{equation}
In this case, the sources in
Eq.~(\ref{eq:amplitudes_luminescence_shift}) can also be classified
according to their symmetry. The amplitudes $F_{1,2}$ and
$\varphi_{1,2}$ have the meaning of projections onto the symmetric
and antisymmetric solutions of Eq.~(\ref{eq:pure_PC_equation}).
Respectively, only symmetric and antisymmetric parts of the sources
contribute into these projections. The exciton radiative decay
contributes only to $\Delta c_2$ because the spatial distribution of
the exciton polarization is determined by the exciton wave function,
which is symmetric by the assumption. The non-resonant polarization
$F$, in turn, does not have to have a definite symmetry and,
therefore, generally contributes to both $\Delta c_1$ and $\Delta
c_2$.  In order to avoid any misunderstanding we have to emphasize,
however, that the functions $h_{1,2}$ do not represent the normal
modes of an infinite photonic crystal. The latter are defined as
solutions of an appropriate boundary problem and generally they do
not have to be even or odd.

%In order to discuss properties of Eq.~(\ref{eq:amplitudes_transfer})
%we consider the important particular case when both the exciton wave
%function and the modulation of the refractive index are symmetric
%with respect to the center of the quantum well, i.e. $\Phi_m(z_m -
%z) = \Phi_m(z_m + z)$ and $n(z_m+z) = n(z_m-z)$. In this case the
%elementary cell of the structure can be chosen to have the explicit
%mirror symmetry with respect to its center (see
%Fig.~\ref{fig:elementary_cell}), and such choice of the elementary
%cell for structures with the mirror symmetry will be implied in what
%follows.

As has been discussed in Ref.~\onlinecite{DispersionMQWPC},
Eqs.~(\ref{eq:amplitudes_transfer}) and
(\ref{eq:amplitudes_luminescence_shift}) do not yet describe the
propagation of the field across an entire elementary cell because
they do not include the transfer across the interface between two
adjacent elementary cells. The problem is that initial conditions
for these solutions are defined at some point inside a given cell,
and in order to use them to describe field in a different cell one
has to introduce the shift of variables $z\rightarrow z\pm nd$,
where $d$ is the period of the structure, and $n$ is the number of
periods separating the two cells. After that one could express the
functions with the shifted arguments as a linear combination of the
original functions, but the most convenient way to describe the
transition from one cell to another is to convert our transfer
matrices to the basis of plane waves. In this basis the field and
its derivative are represented as a superposition of waves
propagating along $z$-axis
\begin{eqnarray}\label{eq:field_plane_waves_definition}
  E &=& E_+ (z)e^{iqz} + E_-(z) e^{-iqz}\\
  dE/dz& =& iqE_+ (z)e^{iqz} -iqE_-(z) e^{-iqz}\nonumber
\end{eqnarray}
where $q = \kappa(z_{\pm})$, and can be naturally presented by a
two-dimensional vector of the form:
\begin{equation}\label{eq:field_mapping}
  \ket{E}= E_+ \ket{+} + E_- \ket{-},
\end{equation}
where
\begin{equation}\nonumber
\ket{+} =\begin{pmatrix}
1\\
0
\end{pmatrix};\hskip 0.6cm
\ket{-} =\begin{pmatrix}
0\\
1
\end{pmatrix}
\end{equation}
are the basis vectors of the respective vector space. More detailed
description of the plane wave representation can be found in
Ref.~\onlinecite{DispersionMQWPC}. The relation between the
coefficients $c_{1,2}$ and the amplitudes $E_{\pm}$ is written as
\begin{equation}\label{eq:relation_basises}
 \begin{pmatrix}
E_+\\
E_-
\end{pmatrix}(z) = M(z) \begin{pmatrix}
c_1\\
c_2
\end{pmatrix}(z),
\end{equation}
where\cite{DispersionMQWPC}
\begin{equation}\label{eq:transition_to_pseudo-waves}
M(z) = \frac{1}{2}
\begin{pmatrix}
 h_{1}(z)+\dfrac{h'_1(z)}{iq} &
h_{2}(z)+\dfrac{h'_2(z)}{iq} \\
 h_{1}(z)-\dfrac{h'_1(z)}{iq} &
h_{2}(z)-\dfrac{h'_2(z)}{iq}
\end{pmatrix}.
\end{equation}
Applying rule (\ref{eq:relation_basises}) to
Eq.~(\ref{eq:amplitudes_transfer}) we obtain
\begin{equation}\label{eq:amplitude_trasfer_waves}
 \ket{E}(z_+) = T \ket{E}(z_-) + \ket{v_m},
\end{equation}
where $T = M(z_+)\widehat{T}_h M^{-1}(z_-)$ is the transfer matrix
through the entire period of the structure in the basis of plane
waves. In the case of structures with the symmetrical elementary
cell this transfer matrix can be presented as\cite{DispersionMQWPC}
\begin{equation}\label{eq:af_representation}
  T =
    \begin{pmatrix}
    af & (\bar{a}f - a\bar{f})/2 \\
    (a\bar{f} - f\bar{a})/2 & \bar{a}\bar{f}
  \end{pmatrix},
\end{equation}
where
\begin{equation}\label{eq:af_s_polarization}
\begin{split}
  a = g_2, \qquad f = g_1 - i S g_2,\\
  \bar{a} = g_2^*, \qquad \bar{f} = g_1^* + i S g_2^*,
\end{split}
\end{equation}
and
\begin{equation}\label{eq:g12_s_polarization}
  \begin{split}
  g_1 = \frac{1}{\sqrt{W_h}}\left[ h_1(z_+) +
  \frac{h_1'(z_+)}{iq}\right], \\
  g_2 = \frac{1}{\sqrt{W_h}}\left[ iq h_2(z_+) +
  h_2'(z_+)\right].
  \end{split}
\end{equation}
The functions $g_{1,2}(\omega)$, which are obviously not unique, are
chosen to make clear the transition to the limiting case of
structures with spatially uniform refractive index. In this case
choosing $h_1(0) = h_2'(0) = 1$ one has $W_h = 1$ and $g_{1,2} =
\exp(i q z_+)$. The function $S(\omega) = -2\pi\omega^2\tilde\chi
\varphi_1^2/(qc^2)$ introduced in Eq.~(\ref{eq:af_s_polarization})
quantifies the interaction of the excitons with light. For the
single-pole form of $\chi(\omega)$ it has the form
\begin{equation}\label{eq:Lorentzian_S_s_polarization}
  S = \frac{\Gamma_0}{\omega - \omega_0 - \Delta\omega +  i\gamma},
\end{equation}
where $\Gamma_0 = 2\pi\alpha\omega^2\varphi_1^2/qc^2$ is the
radiative decay rate. Because of the direct relation between the
functions $\tilde\chi$ and $S$ (they differ only by a factor slowly
changing with frequency) we will for brevity refer to the function
$S(\omega)$ as the exciton susceptibility.

The source term (\ref{eq:amplitudes_luminescence_shift}) in the
basis of plane waves takes the form
\begin{equation}\label{eq:amplitudes_shift_waves}
 \ket{v_m}  = M(z_+)\begin{pmatrix}
\Delta c_1\\
\Delta c_2
\end{pmatrix} = - F_2\frac{4\pi \omega^2 }{c^2} \ket{a} +
 F_1\frac{4\pi \omega^2 }{c^2}\ket{s} -
  \frac{2\tilde\Sigma S q}{\varphi_1} \ket{s},
\end{equation}
where
\begin{equation}\label{eq:a_s_vectors_definitions}
 \ket{a} = \frac{1}{2} \begin{pmatrix}
g_1\\
g^*_1
\end{pmatrix}, \qquad \ket{s} = \frac{1}{2iq} \begin{pmatrix}
g_2\\
-g^*_2
\end{pmatrix}.
\end{equation}
The index $m$ in the notation $\ket{v_m}$ reminds that all the
relevant quantities can depend on the number of the well and should
be taken for a particular well.

\section{Radiative boundary conditions and the field emitted by
an $m$-th well}
 \label{sec:Greens_function}

Equation (\ref{eq:amplitude_trasfer_waves}) expresses the field at
the right boundary of the elementary cell in terms of the field
given at the left boundary. Formally, it can be understood as the
general solution of a Cauchy problem (the general solution of the
homogeneous equation plus a particular solution of the inhomogeneous
one). The amplitudes $E_\pm$ at the left boundary represent two
independent parameters that can be chosen to satisfy any particular
initial or boundary conditions. To represent a radiation coming out
of the structure the field must satisfy radiative boundary
conditions that require that outside the structure there must be
only outgoing waves. Our objective now is, therefore, to find such
$E_\pm$ that would satisfy this condition.To this end we consider an
$N$ layer structure embedded into an environment with the refractive
index $n_{out}$. Scattering of light by the interfaces between the
terminal layers of the structure and the surrounding medium is
described by the matrices
\begin{equation}\label{eq:transfer_interface_scattering}
 T_{L,R} =  \frac{1}{1+\rho_{L,R}}\begin{pmatrix}
    1 & \rho_{L,R} \\
    \rho_{L,R} & 1
  \end{pmatrix},
\end{equation}
where
\begin{equation}\label{eq:interface_Fresnel}
  \rho_{L,R} = \frac{n_{out} \cos\theta_{L,R} - n(z_{L,R})
\cos\theta(z_{L,R})}{n_{out} \cos\theta_{L,R} + n(z_{L,R})
\cos\theta(z_{L,R})}.
\end{equation}
Here $z_{L,R}$ are the coordinates of the left and the right ends of
the structure, respectively. The angle of propagation is determined
by $\tan\theta(z) = k/\kappa(z)$. The outgoing waves propagate at
the angles following from Snell's law $n_{out}\sin\theta_{L,R} =
n(z_{L,R})\sin\theta(z_{L,R})$.

We impose the radiative boundary conditions assuming first that the
sources are localized only in the $m$-th layer. We require that in
the half-spaces $z< z_L$ and $z>z_R$, the field outside the
structure would have the form of the wave propagating respectively
to the left, and to the right. The former field can be described by
a basis vector $\ket{-}$ of the two-dimensional vector space
introduced in Eq.(\ref{eq:field_mapping}): $E=E_-^{(m)}\ket{-}$, and
the later one is proportional to the other basis vector $\ket{+}$:
$E=E_+^{(m)}\ket{+}$. Using the results of the previous Section we
can find the following relation between the fields outside the
structure
\begin{equation}\label{eq:boundary_fields_relation}
  E_+^{(m)}\ket{+} = T_R T(N, m + 1)\ket{v_m} + E_-^{(m)} T_R
T(N, 1) T^{-1}_L\ket{-},
\end{equation}
where
\begin{equation}\label{eq:partial_structure_transfer}
  T(N, m) = T_N \ldots T_m
\end{equation}
is the transfer matrix through the part of the structure obtained as
a product of the transfer matrices through the individual layers.
Eq.~(\ref{eq:boundary_fields_relation}) is obtained by directly
applying Eq.~(\ref{eq:amplitude_trasfer_waves}) to the field at the
left end of the structure. This state is transferred through the
entire structure in a usual way by a simple multiplication of the
transfer matrices describing each period of the structure. This
procedure results in the term proportional to the transfer matrix
$T(N,1)$. Transfer across the luminescent layer results in an
additional contribution as is given by the second term in
Eq.~(\ref{eq:amplitude_trasfer_waves}). After being emitted this
field is then transferred across remaining $N-m$ layers yielding the
term proportional to $T(N, m + 1)$. The matrices $T_{L,R}$ take into
account reflection of the radiation at the interface between the
terminal layers of the structure, and the outside world.

Multiplication of Eq.~(\ref{eq:boundary_fields_relation}) from the
left by $\bra{+}$ and $\bra{-}$ gives the system of two
inhomogeneous equations with respect to $E_\pm^{(m)}$. The solution
of this system is
\begin{equation}\label{eq:field_outside_solution}
\begin{split}
 E_-^{(m)} = & -\frac{\bra{-}T_R
T(N, m + 1)\ket{v_m}}{\bra{-}T_{PC}\ket{-}}, \\
E_+^{(m)} = &\frac{\bra{+}T_L T^{-1}(m, 1)
  \ket{v_m}}{\bra{-}T_{PC}\ket{-}},
\end{split}
\end{equation}
where $T_{PC}=T_R T(N,1) T_L^{-1}$ is the transfer matrix through
the whole structure including the interfaces between the terminating
layers and the surrounding medium.

Equations (\ref{eq:field_outside_solution}) can be used to derive
expression for Green's function defined as a function relating the
radiated field with the source. We can write the amplitudes of the
waves outside in the form
\begin{equation}\label{eq:field_outside_Green}
 E_\pm^{(m)} = \mathcal{G}_\pm^{(s)}(m)\left(F_1 - \tilde\Sigma \varphi_1 \tilde\chi\right)\frac{4\pi\omega^2}{c^2} +
 \mathcal{G}_\pm^{(a)}(m)\left(F_2 - \tilde\Sigma \varphi_2 \tilde\chi\right)\frac{4\pi\omega^2}{c^2},
\end{equation}
where
\begin{equation}\label{eq:Green_functions_definition}
\mathcal{G}_-^{(s,a)}(m) = \mp t_{N}\bra{-}T_R T(N, m + 1)\ket{s,a},
\qquad
\mathcal{G}_+^{(s,a)}(m) = \pm t_{N}\bra{+}T_L T^{-1}(m,1)\ket{s,a}.
\end{equation}
Here we take into account the definition of the transmission
coefficient through the whole structure $t_N$ in terms of the
transfer matrix $t_N=\bra{-}T_{PC}\ket{-}^{-1}$. We would like to
emphasize that these expressions are valid with the vectors
$\ket{s}$ and $\ket{a}$ defined by
Eqs.~(\ref{eq:a_s_vectors_definitions}) even for non-symmetrical
structures where $\ket{s}$ and $\ket{a}$ do not imply symmetrical
properties.

Equation~(\ref{eq:field_outside_solution}) can also be used to find
the distribution of the field created by the source \emph{inside}
the structure. This can be achieved in two different ways. One can
start, for instance, with field $E^{(-)}_m\ket{-}$ on the left-hand
side of the structure and propagate it across using
transfer-matrices. When a luminescent layer is reached,
Eq.~\ref{eq:amplitude_trasfer_waves} should be employed to describe
transfer across it. Alternatively, one can propagate
$E^{(-)}_m\ket{-}$ to find field in the elementary cells to the left
of the luminescent layer, and propagate $E^{(+)}_m\ket{+}$ from the
right to determine field in the cells to the right of it.

We would also like to add that Eq.~(\ref{eq:field_outside_Green})
can be used also for studying the directional and in-plane
distributions of the field. For example, the standard problem of a
point source can also be considered since the in-plane distribution
of the source is taken into account by 2D Fourier transform
(\ref{eq:sources_Fourier}).

Equation~(\ref{eq:field_outside_Green}) shows that the field outside
is determined by two parameters found as convolutions of the sources
with the functions $h_{1,2}(z)$. As has been demonstrated in
Ref.~\onlinecite{DispersionMQWPC}, these functions provide complete
description of photonic modes of a respective infinite system. From
this perspective the result of Eq.~(\ref{eq:field_outside_Green})
might seem expected. Indeed, it \textit{looks} somewhat similar to a
standard construction where the response is determined by a
superposition of modes with amplitudes determined by the projections
of the excitation onto the modes. This analogy, however, is
misleading for the case under consideration. There are several
important features distinguishing Eq.~(\ref{eq:field_outside_Green})
from the standard Green's function formalism. First, as has been
noted, the functions $h_{1,2}(z)$ do not have to coincide with
photonic modes even locally (inside a particular elementary cell).
Second, Eq.~(\ref{eq:field_outside_Green}) is written for a finite
structure when the applicability of the modes of an infinite
structure can not be trivially justified. Third,
Eq.~(\ref{eq:field_outside_Green}) does not require restrictive
properties of the operators governing the propagation of light (say,
hermicity) and, in particular, with slight modifications remains
valid in the presence of losses in the dielectric
($\mathrm{Im}[n(z)]\neq0$) while in this case even the notion of the
projection has to be carefully examined.

Below we will concentrate mostly on the case when the elements of
the structure and the structure itself have the mirror symmetry. As
a result, the index mismatch between the surrounding medium and the
terminal layers is the same for both boundaries, so that one has
$T_L=T_R=T_\rho$. It is interesting to note that in such structures
the solutions given by Eq.~(\ref{eq:field_outside_Green}) do not
necessarily guarantee the symmetry of the radiation emitted to the
left and to the right of the structure. For such symmetry to take
place one has to prove that
\begin{equation}\label{eq:actual_symmetry_condition}
 E_-^{(N-m+1)}=E_+^{(m)}.
\end{equation}
This relation can indeed be proven but only in the case when both
the structure and the sources are symmetrical with respect to the
centers of the elementary cells. While the exciton related resonance
source term does have the required symmetry, the non-resonance
contribution is not necessarily symmetric if  both  $F_{1,2}\ne0$
(see Eq.~(\ref{eq:amplitudes_shift_waves}) and
Eq.~(\ref{eq:field_scalar_equation})). If this is indeed the case
than  radiation emitted to the right would not have the same
characteristics as radiation emitted to the left. When $F_2 \equiv
0$, however, Eq.~(\ref{eq:actual_symmetry_condition}) can be proven
with the help of relation $T^{-1}\ket{s} = - \sigma_x \ket{s}$,
where $T$ is defined in Eq.~(\ref{eq:af_representation}), and
$\sigma_x$ is the standard Pauli matrix.

As follows from Eq.~(\ref{eq:field_outside_Green}) the field
radiated due to the exciton recombination has, as expected, the
resonant character. Thus, in resonant PCs the recombination
determines the spectrum of the radiated field at frequencies close
to $\omega_0$, while the non-resonant sources specified by $F(z)$
are responsible for the background component characterized by a
relatively smooth frequency dependence. This contribution may become
important farther away from the exciton frequency. For the problem
of the exciton luminescence in the resonant PCs these non-resonant
sources are not important and, therefore, below we assume that the
only source of radiation is the exciton recombination and neglect
the non-resonant contribution. The expression for the radiated field
essentially simplifies in this case and takes the form
\begin{equation}\label{eq:emission_outside_Green}
 E_\pm^{(m)} = -\mathcal{G}_\pm(m)\Sigma_m \frac{2 S_m q}{{\varphi_m}_1},
\end{equation}
where we explicitly show the dependence of the source on the number
of the layer and
\begin{equation}\label{eq:symmetric_Green}
 \mathcal{G}_\pm(m) = \pm t_N\bra{\pm} T_{\mp}(m)T_\rho\ket{s}.
\end{equation}
Here we have taken into account that $\tilde\Sigma_m = \Sigma_m$
when $F\equiv0$ and have dropped the superscript $(s)$ since the
only relevant Green's function in this case is
$\mathcal{G}_\pm^{(s)}(m)$. We have also introduced partial transfer
matrices $T_-(m)=T_\rho T^{-1}(m, 1)T_\rho^{-1}$ and $T_+(m)=T_\rho
T(N, m + 1)T_\rho^{-1}$, which have the property $T_-^{-1}(m)T_+(m)
= T_{PC}$.

\section{The luminescence spectrum of resonant photonic crystals}
 \label{sec:Emission_spectrum}
\subsection{Quasi-classical approach to luminescence}
In this section we will apply general results of previous sections
to the problem of the luminescence spectrum of multiple-quantum-well
based resonant photonic crystals. Luminescence is one of the
manifestations of spontaneous emission, and as such is a purely
quantum electrodynamic phenomenon. At the same time, the
non-coherent radiation produced due to luminescence in many
situations can still be described as classical electromagnetic field
with randomly changing amplitude and phase. The average value of
such a field is equal to zero, while its variance, defined as
$\langle E^2({\bf r})\rangle$, is identified with intensity, where
angular brackets $\langle \cdots\rangle$ symbolize averaging over
appropriately defined distribution function. Such a quasi-classical
approach to spontaneous emission has been used previously in a great
number of different situations. Just as a small sample of papers
dealing with applications of classical electrodynamics to the
problem of spontaneous emission we can refer to Ref.
\onlinecite{Morawitz:1969,Dowling:1992,Cavity_Luminescence:1996,Dodd:1997,Yariv:2000}.

One of the methods of reproduction of non-coherent classical field
is a Langevin-like approach, in which the polarization sources
appearing in macroscopical Maxwell equations
(\ref{eq:Mawell_equation_start}) are considered as random functions
of time and coordinates whose statistical characteristics should be
determined either from experiment or from fully quantum
microscopical theory. In what follows we will neglect the
non-resonant contribution to polarization $F$, and characterize
statistical properties of the resonant source function
$\Sigma_m(\mathbf{\rho},t)$ by a correlation function
\begin{equation}\label{eq:correl_funct}
\langle\Sigma_{i,m}(\mathbf{\rho},t)\Sigma_{j,l}(\mathbf{\rho}^\prime,t^\prime)\rangle
=
\delta_{ij}\delta_{ml}K_\Sigma\left(\mathbf{\rho}-\mathbf{\rho^\prime},t-t^\prime\right),
\end{equation}
where  $\langle \cdots\rangle$ signifies statistical averaging over
various realizations of the non-coherent exciton polarization,
indexes $i$ and $j$ designate Cartesian coordinates in the
$(x,y)$-plane, and $m$ and $l$ are well numbers.
Equation~(\ref{eq:spectral_density}) implies that the fluctuations
of the non-coherent exciton polarization are (i) statistically
uniform in time and space,   (ii) direction of the non-coherent
polarization is distributed isotropically in the plane of the
structure, with various components of the polarization vector
independent of each other, and (iii) source functions in different
wells do not correlate with each other.

This correlation function can only be found from a microscopical
theory of electron-hole relaxation processes. An example of such a
theory, which provide a more quantitative justification for our
phenomenological approach and shows relation of this correlation
function to microscopic characteristics of quantum wells, is
presented in Appendix. These calculations demonstrate that all three
assumptions regarding properties of the source correlations are
justified. The most important of them is the assumption of
independence of the source functions in different wells.  We would
like to emphasize here that this assumption \emph{does not mean}
that we neglect radiative coupling between wells. As it is explained
in Appendix the source term $\Sigma_m(\mathbf{\rho},t)$ is
determined by excitons populating non-radiative states so that the
electromagnetic field associated with them decays exponentially
outside of the wells and does not contribute to the radiative
coupling. This coupling is described by the common radiative field
$E$ which is created by and act on all quantum wells in the system
as it was explained in previous section
\ref{sec:Macroscopic_Maxwell_equations}. It is interesting to note
that Maxwell equation (\ref{eq:Mawell_equation_start}) describes
this coupling even though the average value of the non-coherent
field is equal to zero. The correlation function for the fourier
transformed source function $\Sigma_m(\omega,{\bf k})$ is given by
spectral density $\Xi_m(\omega,k)$ defined as
\begin{equation}\label{eq:spectral_density}
\langle\Sigma_m(\mathbf{k},\omega)\Sigma_l(\mathbf{k}^\prime,\omega^\prime)\rangle
=\Xi_m(\omega,k)\delta(\omega-\omega^\prime)\delta(\mathbf{k}-\mathbf{k}^\prime)\delta_{ml}
\end{equation}
which is a Fourier transform of the time-position correlation
function given in Eq.(\ref{eq:correl_funct}).

The field created by such source is characterized by a spectral
intensity $\mathcal{I}(\mathbf{k},\omega)$ defined as
\begin{equation}\label{eq:intentisty}
\langle E(\mathbf{k}_1,\omega_1) E(\mathbf{k}_2,\omega_2)\rangle =
\mathcal{I}(\mathbf{k}_1,\omega_1)\delta(\mathbf{k}_1 -
\mathbf{k}_2)\delta(\mathbf{\omega}_1 - \mathbf{\omega}_2).
\end{equation}
Applying Eq.~(\ref{eq:spectral_density}) to this equation we find
the spectral intensity of radiation emitted by the entire structure
in the form
\begin{equation}\label{eq:field_spectral_density}
\mathcal{I}_\pm(\omega,k) = 4\,\sum_m\Xi_m(\omega,k)
|\mathcal{G}_\pm(m; \omega,k )|^2 \left|\frac{qS_m(\omega,
k)}{{\varphi_m}_1}\right|^2.
\end{equation}
which implies that the field emitted by different wells adds in a
non-coherent way. This general expression allows analyzing both the
frequency and the directional dependence of the luminescence
spectrum. In this work we restrict our consideration to the  waves
emitted along the growth direction of the structure (i.e. $k=0$).
The directional distribution of the radiation will be studied
elsewhere.

Equation~(\ref{eq:field_spectral_density}) shows that the form of
the luminescence spectrum is determined by several factors with
different frequency dependencies. The exciton susceptibility,
$S(\omega)$ for instance,  strongly reduces the luminescence far
away from the exciton frequency $\omega_0$. The spectral density
$\Xi(\omega)$ is expected to show a weak frequency dependence at the
scale of the width of the polariton stop-band. The factor
$|\mathcal{G}(m; \omega,k )|^2$, according to
Eq.~(\ref{eq:symmetric_Green}), is the product of two terms. One is
the transmission coefficient $t_N$, which may have strong frequency
dependence following the singularities at the eigenfrequencies of
the quasi-modes of the structure. These singularities determine the
fine structure of the luminescence spectrum. The second term is
responsible for the variations in the luminescence intensity at a
much larger scale. Equation~(\ref{eq:field_spectral_density}) can be
used to analyze spectra of luminescence of various types of MQW
structures. Several examples are presented in the subsequent parts
of this section.

\subsection{The luminescence spectrum of finite periodic structures}
\subsubsection{General expression for intensity of emission}
We first apply our general results to  the  case  structures built
of identical layers, so that ${\varphi_m}_1 = \varphi_1$ and $S_m =
S$, i.e. all quantum wells are characterized by the same exciton
frequency $\omega_0$. We would like to note that in order to have
the structure with the mirror symmetry it must contain an integer
number of the elementary cells. This means, in particular, that the
terminating layers are half-barriers. As has been noticed, in this
case the luminescence spectrum is the same at the both sides of the
structure. For the concreteness, we will consider the field radiated
to the right (i.e. $E_+$).

In the symmetric case we can perform the summation over the quantum
wells using the fact that all partial transfer matrices posses the
mirror symmetry and can be presented in the
form\cite{Optical_Properties_MQWPC}
\begin{equation}\label{eq:theta_beta_representation}
T(\theta, \beta) = \begin{pmatrix}
    \cos \theta - i\sin\theta \cosh\beta & - i\sin\theta
\sinh\beta \\
    i\sin\theta \sinh\beta & \cos \theta + i\sin\theta \cosh\beta
  \end{pmatrix},
\end{equation}
where
\begin{equation}\label{eq:theta_beta_a_f}
  \cos\theta = \frac 1 2 (af + \bar{a}\bar{f}), \qquad
  \tanh \beta = \frac{\bar{a}f - a\bar{f}}{af - \bar{a}\bar{f}}.
\end{equation}
The parameter $\theta$ determines the polariton spectrum of an
infinite structure and is defined as $\theta = Kd$, where $K$ is the
polariton Bloch wave-number.

The representation (\ref{eq:theta_beta_representation}) is
convenient because all $T_-(m)$ are characterized by the same
$\beta$, while the spectral parameter of $T_-(m)$ is merely
$-m\theta$. Thus, $T_-(m)$ can be written as
\begin{equation}\label{eq:T_-_almost_spectral_representation}
\begin{split}
  T_-(m) = & \,e^{i m \theta} U(\beta/2)\ket{+}\bra{+}
  U^{-1}(\beta/2)\\
   & + e^{-i m \theta} U(\beta/2)\ket{-}\bra{-}
U^{-1}(\beta/2),
\end{split}
\end{equation}
where $U(\beta/2)\equiv T_{\rho}T_H(\beta/2)$, and $T_H$ is a matrix
describing a hyperbolic rotation with a dilation
\begin{equation}\label{eq:T_H_psi_definition}
  T_H(\beta) =
 e^{\beta} \begin{pmatrix}
   \cosh\beta & -\sinh\beta \\
   -\sinh\beta & \cosh\beta
 \end{pmatrix}.
\end{equation}
Matrix $T_{\rho}$ takes into account reflection and transmission at
the external boundaries of the system.

Using Eq.~(\ref{eq:T_-_almost_spectral_representation}) one can find
\begin{equation}\label{eq:sum_G_m_calculated}
 \frac{4q^2}{|t_N|^2}\sum_m  \left|\mathcal{G}_+(m)\right|^2 =
 \frac{\sinh N\theta''}{\sinh\theta''}
  \left[|A|^2 e^{-(N+1)\theta''} + |B|^2
e^{(N+1)\theta''} \right]
  + \frac{\sin N\theta'}{\sin \theta'}\left[AB^*
e^{i\theta'(N+1)}+ A^*B e^{-i\theta'(N+1)}\right],
\end{equation}
where we have introduced real and imaginary parts of the
dimensionless Bloch number, $\theta$: $\theta = \theta' + i\theta''$
and parameters
\begin{equation}\label{eq:A_B_definitions}
\begin{split}
 A = & \frac{1}{1+\rho}\left[ (g_2- \rho g^*_2) \cosh^2(\beta/2) -
 \frac 1 2 (g_2^* - \rho g_2)\sinh \beta\right]\\
 B = -&\frac{1}{1+\rho} \left[ (g_2 - \rho g^*_2) \sinh^2(\beta/2) -
 \frac 1 2(g_2^* - \rho g_2)\sinh\beta\right].
\end{split}
\end{equation}

For the purposes of numerical calculations instead of direct
calculations of the parameter $\beta$ it is more convenient to
multiply both parts of Eqs.~(\ref{eq:A_B_definitions}) by
$\sin\theta$ and, then, use
Eq.~(\ref{eq:theta_beta_representation}) to establish direct
relation of corresponding terms with the elements of the transfer
matrix through the period of the structure.
\begin{figure}
  \includegraphics[width=3in]{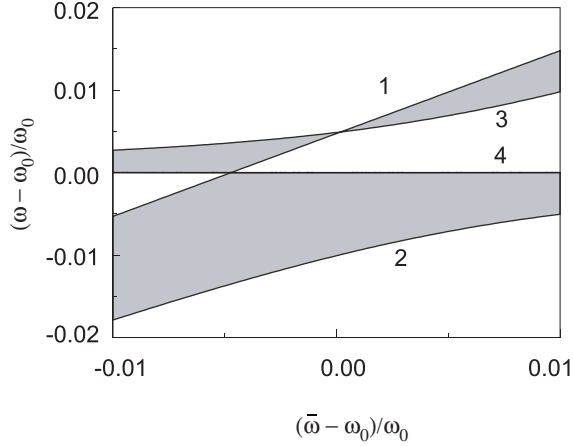}
  \caption{(color online)
 Evolution of polariton band boundaries with detuning of the structure with piecewise modulation of the refractive
 index from exact Bragg condition. The horizontal axis describes the detuning in terms of parameter $\bar{\omega}-\omega_0$,
 where $\bar{\omega}=\pi c/(n_wd_w+n_bd_b)$, $n_w$, $n_b$, $d_w$, $d_b$ and refractive indexes and thicknesses of well and barrier layers respectively.
 The vertical axis shows a relative difference between frequency $\omega$, and exciton frequency $\omega_0$.}\label{fig:band_structure}
\end{figure}
%It is immediately seen from Eq.~(\ref{eq:sum_G_m_calculated}) that
%the luminescence is reduced at frequencies corresponding to an
%interior of a forbidden gap. At these frequencies the main
%contribution to the right-hand-side of
%Eq.~(\ref{eq:sum_G_m_calculated}) is due to the exponentially large
%terms $\propto \exp(N \theta'')$. However, these terms are cancelled
%by the exponentially small transmission at these frequencies. As a
%result, $\mathcal{I}(\omega)$ tends to values which are independent
%on the length of the structure. Thus, for sufficiently long
%structures where the polariton band structure essentially effects
%their optical properties the luminescence at forbidden frequencies
%is relatively small, since only wells within the attenuation

While Eq.~(\ref{eq:sum_G_m_calculated}) describes luminescence of a
periodic MQW structure with an arbitrary period, we shall focus our
attention to the most interesting cases of Bragg and near-Bragg
structures,\cite{IvchenkoSpectrum,IvchenkoMQW} in which effects of
periodic modulation of the refractive index and light-exciton
coupling are most pronounced. As we already mentioned in
Introduction the period, $d$, of such structures satisfies a special
resonance condition $\omega_0=\omega_B(d)$, where the exact value of
the resonant frequency $\omega_B$ in systems with periodically
modulated refractive index depends not only on the period of the
structure, but also on details of the
modulation.\cite{JointComplex,DispersionMQWPC} For concreteness we
will assume that the dielectric function reaches its maximum value
at the quantum well and monotonously decreases towards the
boundaries of the elementary cell. In this case the Bragg resonance
takes place when the exciton frequency coincides with the
high-frequency boundary of the photonic band gap, $\Omega_+$. Since
the details of the emission spectrum are determined to a large
extent by the electromagnetic band structure of the systems under
consideration, it is useful to remind the main features of this
structure, which was analyzed in details in a number of
papers.\cite{IvchenkoMQW,IvchenkoSpectrum,DLPRBSpectrum,JointComplex,DispersionMQWPC}
Fig.~\ref{fig:band_structure} shows the dependence of band
boundaries of MQW structure upon its period, where shaded regions
correspond to polariton stop-bands. One can see that at a certain
value of $\bar{\omega}=\pi c/(n_wd_w+n_bd_b)$, where $n_w$, $n_b$,
$d_w$, $d_b$ and refractive indexes and thicknesses of well and
barrier layers respectively, two stop-bands connect at the exciton
frequency $\omega_0$ forming a single wide band-gap. This is the
point of the Bragg resonance, when exciton frequency falls inside a
stop-band of the spectrum, whose width can be much larger than the
width of the exciton resonance.  (The second occurrence of a single
band situation at larger values of $\bar{\omega}$ results from
collapse of one of the gaps, and is a result of random degeneracy
between two exciton polariton branches.) Spectrum of structures only
slightly detuned from the Bragg condition (we will use term
quasi-Bragg for such structures), is characterized by emergence of a
propagating band between the two stop-band. The exciton frequency in
this case belongs to the boundary of the propagating band, which is
situated asymmetrically with respect to the outer band boundaries:
for negative detunings $\omega_0$ is closer to the upper boundary,
while for positive detuning the lower polariton branch eventually
moves closer to it.
%The exciton luminescence also depends on the relation .

Equation( \ref{eq:sum_G_m_calculated}) demonstrates how the
polariton band structure affects the luminescence of the system
under consideration. One can see from this equation that the
structure of the spectrum is characterized by two scales of
frequencies. On a smaller scale the modulations of the intensity of
emission are determined by $|t_N|^2$ term in
Eq.~(\ref{eq:sum_G_m_calculated}), whose maxima correspond to the
real parts of the polariton eigenfrequencies. The modulation of
intensity on this scale depends on the number of periods in the
structure and occurs over frequency intervals of the order of
$v_g/(dN)$, where $v_g$ is the group velocity of the polariton
excitations and $d$ is the period of the structure. The polariton
band structure affects the luminescence on a much larger spectral
scale through the combination of the exciton susceptibility
$S(\omega)$ and the imaginary part of the polariton Bloch number
presented by parameter $\theta^{''}$.

Homogeneous and inhomogeneous broadenings significantly effect the
short-scale modulations of the luminescence allowing for their
observation only in high quality samples at very low temperatures.
At higher temperatures the long scale variations of intensity, which
depend significantly on relations between $\omega_0$ and $\omega_B$,
become predominant. In the case of Bragg structures, when $\omega_0$
is very close to $\omega_B$, Eq.~(\ref{eq:sum_G_m_calculated})
predicts that the luminescence spectrum is mostly concentrated
outside of the polariton stop-band near the edges of the bands of
the exciton polaritons. Indeed, at frequencies inside the forbidden
gap the contribution to $\mathcal{I}(\omega)$ of the exponentially
large terms in the right-hand-side of
Eq.~(\ref{eq:sum_G_m_calculated}) is canceled by the exponentially
small transmission at these frequencies. As a result, only wells
within the attenuation length from the boundaries contribute to the
radiated field. Besides at frequencies far away from the $\omega_0$
the luminescence is subdued by the smallness of the exciton
susceptibility. These qualitative arguments can be supported by
direct calculation of the emission intensity in the neighborhood of
the band edges, where Eq.~(\ref{eq:sum_G_m_calculated}) can be
simplified. In this spectral region  we can represent the spectral
parameter as $\theta = \pi + i \epsilon$ and assume that $\epsilon$
is sufficiently small, so that $N|\epsilon| \ll 1$. Obviously, this
approximation covers a substantial interval of frequencies only for
not very long structures, but it is quite sufficient for a
qualitative analysis of realistic structures.
\begin{figure}
  \includegraphics[width=5.3in]{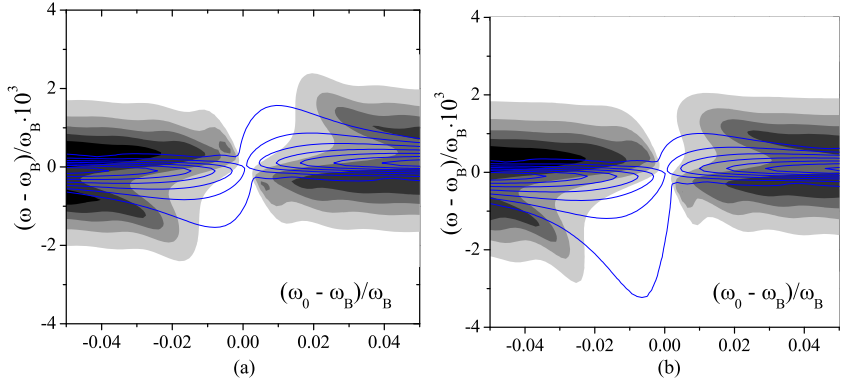}
  \caption{(color online)
  The luminescence spectrum in comparison with a polariton band structure
is shown for quasi-Bragg structures consisting of $60$ layers near
the boundary of the first Brillouin zone. Smooth filling corresponds
to the luminescence calculated using $\Xi =
 1$. The lines are the contour plot of $\theta''$, where the outermost curve
correspond to its smallest value. Frequency changes along the
vertical axis, while, the horizontal one presents detuning from the
Bragg resonance. (a) Pure MQW structure with parameters typical for
Al${}_x$Ga${}_{1-x}$As/GaAs structures: $\Gamma_0 = 75$ $\mu$eV,
$\omega_0 = 1.489$ eV, $\gamma = 300$ $\mu$eV. (b) An example of MQW
based photonic crystal. The exciton related parameters are the same
as in (a). The modulation of the index of refraction is taken to be
$n(z) = 3.4 + 0.1 \cos^{20}(\pi
z/2d)$.}\label{fig:luminescence_band_structure}
\end{figure}

Expending  Eq.~(\ref{eq:field_spectral_density}) in a power series
with respect to the small parameter $\epsilon N$, we can approximate
it by the following simple expression:
\begin{equation}\label{eq:intensity_Bragg_near_edges}
\mathcal{I}(\omega) \approx N \Xi(\omega)|t_N|^2
\left|\frac{S(\omega)}{\varphi_1}\right|^2 \frac{h^2_2q^2}{W_h}.
\end{equation}
An important result immediately demonstrated by this equation is a
linear increase of the intensity of the emission with the number of
quantum wells. This is an expected behavior because of the
transparency of the structure at these frequencies and the
independence of the contributions of different wells to the emitted
light. Another important conclusion following from
Eq.~(\ref{eq:intensity_Bragg_near_edges}) is a relative weakness of
the luminescence of the Bragg structures. The cause of the decrease
in the emission is related to the presence of a broad polariton
stop-band in such structures whose width, $\Delta$, given by
expression $\Delta = \sqrt{2\Gamma_0\omega_0/\pi}$, is much larger
than the width of the exciton susceptibility $S(\omega)$ determined
by non-radiative decay, $\gamma$. In the interior of the stop-band
the luminescence is suppressed by small transmissivity of the
structure in the vicinity of the exciton resonance, while at the
edges of the stop-band it is reduced by the factor of
$\Gamma_0/\omega_0\ll 1$ because of the separation of the band
boundaries from the exciton frequency. The found decrease in
luminescence for Bragg structures is equivalent to the so called
sub-radiance effect obtained in Ref.~\onlinecite{KIRA:2006} on the
basis on fully quantum calculations. We demonstrate here that this
effect has a purely classical origin and is caused by formation of
polariton stop-band in Bragg MQW systems.

Detuning from the Bragg resonance opens up a transparency window
inside the stop-band with exciton frequency coinciding  with one of
the band boundaries (Fig.\ref{fig:band_structure}). On the base of
the same arguments as above one can expect that the emission
spectrum is characterized by two maxima: the stronger one in the
vicinity of the boundary of the propagating band adjacent to
$\omega_0$, and the second, weaker, maximum at the boundary of the
outer polariton band, which is closer to the exciton frequency. The
second outer boundary of the polariton band is so remote from the
exciton frequency that its contribution to emission can be
neglected. Numerical calculations carried out with the exact form of
$\mathcal{I}(\omega)$ confirm these conclusions. The results of
these calculations are presented in
Fig.~\ref{fig:luminescence_band_structure}, where we show the
dependence of the intensity upon frequency and the period of the
structure. The intensity is shown by the shading on the graphs ---
the darker shading corresponds to higher emission. In order to
facilitate a better understanding of the role of the refractive
index contrast we simulated two types of structures: one with a
realistic changes in the refractive index between wells and
barriers, and the other, in which refractive index was assumed
constant throughout a structure. The latter structures are often
called optic lattices because all the modifications in their optical
properties come from the radiative coupling between quantum well
excitons. It is interesting to see a significant difference between
the luminescence spectra of MQW optical lattices and MQW based
photonic crystals. The latter is  asymmetric with respect to the
point of the Bragg resonance, while the former shows complete
symmetry. This feature is clearly related to the asymmetrical
structure of the polariton band gap in structures with modulated
refraction index.\cite{DispersionMQWPC}
\begin{figure}
  \includegraphics[width=5in]{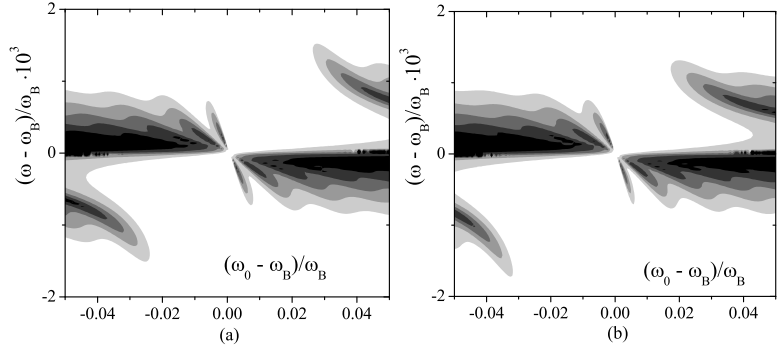}\\
  \caption{The fine structure of the luminescence spectrum.
  The parameters of the structures are the same as in Fig.~\ref{fig:luminescence_band_structure}
 except $\gamma = 30$ $\mu$eV. For better visibility the shadow intensity
 is chosen according to the log-scale. (a) The MQW structure
 with a homogeneous dielectric function. (b) The MQW based photonic
 crystal.
}\label{fig:luminescence_band_structure_fine}
\end{figure}
In order to emphasize the relationship between the luminescence
spectrum and polariton band structure, the spectrum in this figure
is presented together with the polariton stop band. The latter is
shown with the help of level curves of the imaginary part of the
dimensionless Bloch vector $\theta''$. In an ideal system without
any broadenings, $\theta''$ would have been zero everywhere outside
of the stop-band. In real systems, of course, $\theta''$ is not zero
everywhere because of the exciton broadening. This  makes the notion
of the stop-band not very well defined, and, in particular, the
edges of the gap can not be determined unambiguously. However, at
the frequency, which would correspond to the band edge in a system
without broadening, the imaginary part of the polariton Bloch
wave-number drastically increases. This increase can be traced on
the level curves of $\theta''$ in
Fig.~\ref{fig:luminescence_band_structure}, where outer curves
correspond to the smallest value of $\theta''$. It is seen that the
maxima of the luminescence spectrum approximately follow these lines
when the relation between the the exciton frequency and the period
of the structure changes. The exact position of the maxima is
determined by an interplay between a smaller value of $\theta''$
(and, hence, a higher transmission) and a smaller distance from the
exciton frequency (a higher value of $S(\omega)$). Comparing the
spectrum shown in Fig.~\ref{fig:luminescence_band_structure} with
the band structure shown in Fig.~\ref{fig:band_structure} one has to
notice the different frequency scales of these figures. The
frequency region covered in
Fig.~\ref{fig:luminescence_band_structure} includes only the small
transparency window around the exciton frequency and only the
closest to it outer polariton band.

\begin{figure}
  \includegraphics[width=4in]{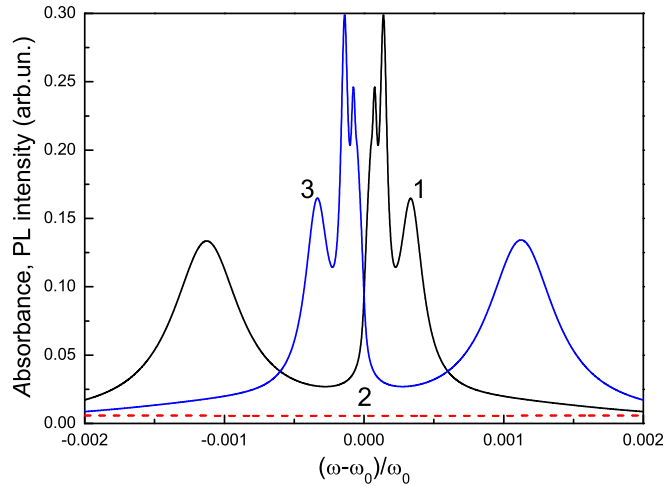}\\
\caption{(color online) Fine structure of the PL spectra for the 100
wells MQW structures with different periods. Curves 1, 2 and 3 are
calculated for the period respectively smaller, equal, and larger
than the Bragg resonance value. The indexes of refraction of wells
and barriers in this calculation were assumed equal to each
other.}\label{fig:intensity_versus_frequency}
\end{figure}
For sufficiently smaller value of the exciton broadening the fine
structure becomes clearly visible as is seen in
Fig.~\ref{fig:luminescence_band_structure_fine}. As we mentioned
above the maxima of the luminescence forming this fine structure
result from the periodic dependence of the transmission on
frequency. These maxima appear as the characteristic scars on the
spectrum presented in this figure. More clear representation of
these features of the luminescence spectrum can be given by direct
plotting of the intensity as function of frequency for different
values of the de-tuning of the structure from the Bragg resonance,
as shown in Fig.~\ref{fig:intensity_versus_frequency}, which was
obtained neglecting the modulation of the refractive index.
\begin{figure}
  \includegraphics[width=4in]{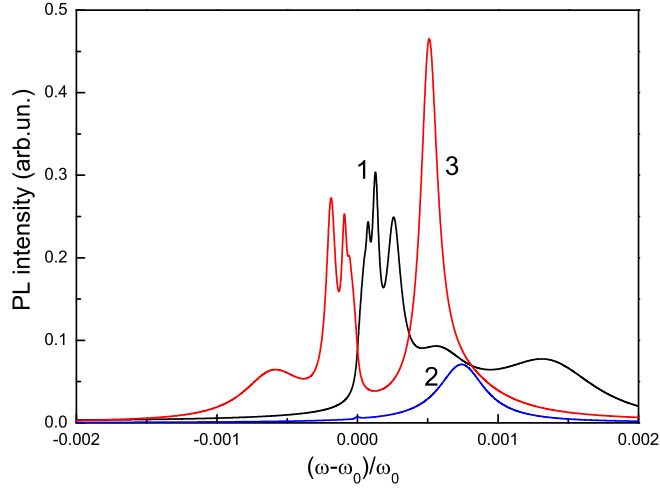}\\
\caption{(color online)The same spectra as in Fig.
\ref{fig:intensity_versus_frequency} but with cladding layer of
thickness
$d_c=d_b+d_w/2$}\label{fig:intensity_versus_frequency_cladding}
\end{figure}
\subsubsection{Comparison with experiment and the role of
inhomogeneous broadening} Comparing our calculations with
experimentally observed spectra,\cite{HUBNER:1999,MINTSEV:2002} one
should take a few considerations into account. First of all, the
direct quantitative comparison is rather difficult because the
experimental spectra are influenced by details of the entire
experimental sample, and not just by its MQW part. For instance, the
details of the cladding layer can significantly influence the
observed luminescence spectrum. In order to illustrate this point we
used the general formulas derived in the paper to calculate the
emission intensity in the presence of the cladding layer. The
results of these calculations are shown in
Fig.~\ref{fig:intensity_versus_frequency_cladding}, where one can
notice significant changes introduced by the cladding layer to the
spectra. Second, in photoluminescence experiments with long MQW
structures the intensity of pump radiation is not uniform along the
structure, which results in different source functions for different
wells. This circumstance also affects observed spectra  as can be
demonstrated by direct computations using general formulas obtained
\begin{figure}
  \includegraphics[width=5in]{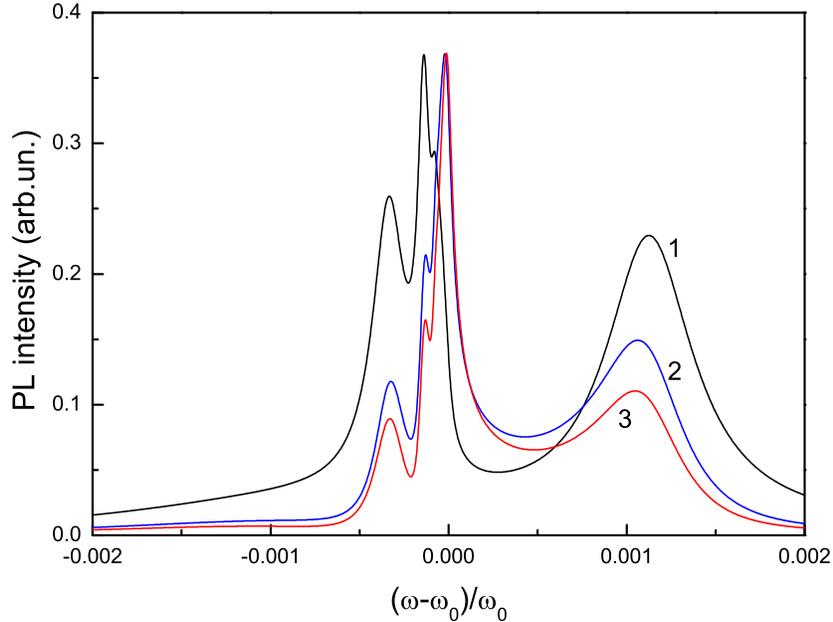}
  \caption{(color online) The luminescence spectrum of the almost Bragg $100$ layers structure with exponentially decaying source
  term characterized by the decay rate $\alpha$: curve 1 corresponds
  to  $\alpha=0$, curve 2 to $\alpha=0.1$  and curve 3 represents $\alpha=0.2$}\label{fig:pump_absorp}
\end{figure}
in this Section. To this end we assumed that the source function,
$\Xi_m(\omega)$, which appears in
Eq.~(\ref{eq:field_spectral_density}) can be presented as an
exponentially decreasing function of the well number, $m$:
$\Xi_m(\omega)\propto \exp({-\alpha md})$, where parameter $\alpha$
represents an inverse attenuation length of the pump. Using this
representation for the source function in
Eq.~(\ref{eq:field_spectral_density}) we numerically calculated
emission intensity with different values of parameter $\alpha$. The
results of these calculations are shown in
Fig.~\ref{fig:pump_absorp}, where luminescence spectra with
$\alpha=0$ and $\alpha=0.2$ are compared. This figure clearly
demonstrates that inhomogeneity of the source function can have a
significant impact on the observed spectra.

Having in mind mentioned circumstances, we will not attempt to
quantitatively reproduce experimental spectra, focusing instead on
the most significant features, which most likely have intrinsic
origin. Comparing experimental results of
Refs.~\onlinecite{HUBNER:1999} and \onlinecite{MINTSEV:2002}, one
can notice that despite of quantitative difference between these two
experimental spectra, they share one common feature, which is, at
the same time is in a striking contrast with results of our
calculations. According to our predictions, the luminescence must be
most intense in the vicinity of the exciton frequency, while the
experiments show that out of two most pronounced maxima of the
emission, the one, which is farther away from $\omega_0$ is
brighter.
\begin{figure}
  \includegraphics[width=4in]{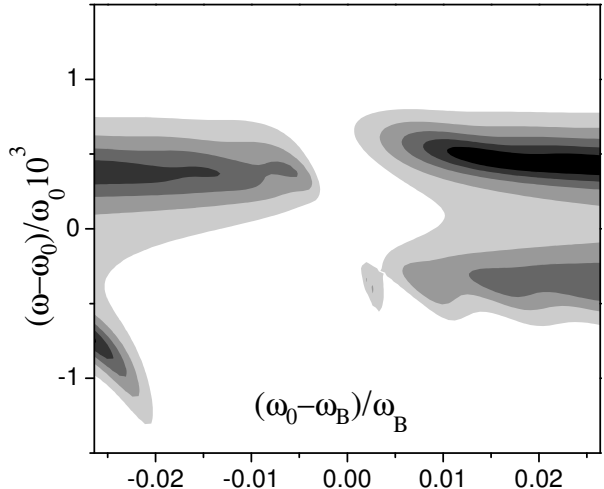}\\
\caption{The same spectra as in
Fig.~\ref{fig:luminescence_band_structure}a, but with inhomogeneous
broadening taken into account within the effective medium
approximation. The parameter of inhomogeneous broadening was chosen
to be equal to $\sigma=200\hspace {2 pt} \mu eV$.}
\label{fig:lumin_profile_inhom_broad}
\end{figure}
\begin{figure}
  \includegraphics[width=4in]{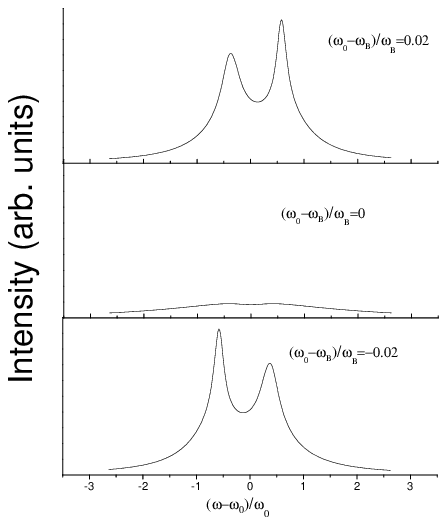}\\
\caption{Cross-sections of the intensity profile shown in Fig.
\ref{fig:lumin_profile_inhom_broad} for three different values of
detuning from exact Bragg
resonance.}\label{fig:intensity_versus_frequency_inhom_broad}
\end{figure}
One probable reason for this discrepancy is the inhomogeneous
broadening of excitons, which has not yet been taken into account in
our calculations. In this work we include effects due to the
inhomogeneous broadening into consideration using a simple model of
effective medium.\cite{AndreaniKavokinIB,{OmegaDefectPRB}} Within
this model one neglects spatial dispersion of excitons and assumes
that inhomogeneous broadening is caused by spatial fluctuations of
exciton frequency $\omega_0$. It is further assumed that these
fluctuations can approximately be taken into account by replacing
exciton susceptibility, Eq.~(\ref{eq:exciton_susceptibility}), in
all relevant equations, with its average value
\begin{equation}\label{eq:effect_susc}
\chi(\omega)_{eff}=\int{\frac{\alpha}{z-\omega-i\gamma}\rho(z)dz},
\end{equation}
where $\rho(z)$ is a distribution function of exciton frequencies.
In the case of a not very strong inhomogeneous broadening this
function can be approximated by a Gaussian:
\begin{equation}\label{eq:distr_func}
\rho(z)=\frac{1}{\sqrt{\pi}\sigma}\exp{\left[-\displaystyle{\frac{\left(z-\overline{\omega}_0\right)^2}{\sigma^2}}\right]},
\end{equation}
where $\overline{\omega}_0$ is an average exciton frequency, and its
r.m.s. value, $\sigma$ determines the width of the distribution.
This model of the inhomogeneous broadening was first introduced in
Ref.~\onlinecite{AndreaniKavokinIB} on heuristic basis for
calculations of reflection and transmission spectra of MQW
structures, and later more rigorously justified in
Ref.~\onlinecite{OmegaDefectPRB}. While derivation of this model
carried out in Ref.\onlinecite{OmegaDefectPRB} cannot be directly
applied to the luminescence problem, we still believe on the ground
of physical arguments similar to those put forward in
Ref.~\onlinecite{AndreaniKavokinIB}, that the effective medium
approach can give qualitatively accurate description of the role of
inhomogeneous broadening in luminescence spectra, at least, for
emission in directions close to normal. Indeed, light emitted by
quantum well excitons in the close to normal direction leaves the
quantum well without moving significantly in the in-plane direction,
and thus without experiencing significant scattering due to in-plane
disorder. Accordingly, the main effects of the in-plane disorder in
this case is that excitons localized in different regions of the
sample emit light at different frequencies. A probe with a
sufficiently large aperture (a typical situation in luminescence
experiments unless one deals with microluminescent spectra) would
collect light emitted by all these excitons, effectively averaging
out their susceptibility.

The result of numerical computation of emission spectra with
Eq.~(\ref{eq:effect_susc}) for exciton susceptibility are presented
in Figs.~(\ref{fig:lumin_profile_inhom_broad}) and
(\ref{fig:intensity_versus_frequency_inhom_broad}). The first of
these figures show that taking into account the inhomogeneous
broadening resulted in spectral redistribution of the emission
intensity from peaks closer to the central exciton frequency to
those that are farther away from it. The latter are now brighter
than the former in a qualitative agreement with experimental
spectra. This point is demonstrated even more clear in the second of
these figures, which shows emission spectra for three values of
detuning from Bragg resonance. Qualitatively this affect can be
understood by noticing that by averaging the exciton susceptibility
we essentially smoothed its resonance dependence on the frequency
reducing, therefore, effect of decreasing susceptibility on the
emission intensity. In this situation the intensities of
luminescence peaks are determined by an interplay between affects
due susceptibility and transmissivity of the structure. These
calculations show that inhomogeneous broadening can be in principle
responsible for observed luminescent spectra, while it is clear that
a quantitative agreement with experiment would require more rigorous
treatment of inhomogeneous broadening as well as taking into account
such effects as inhomogeneity of pump and cladding layers.

\subsection{The luminescence spectrum of structures with defects}
One of the main reasons for the interest to resonance photonic
crystals derives from possibilities to manipulate their optical
properties through modification of their structure. One of the
possible approaches includes intentional violation of the
periodicity of the PCs by introducing one or several defects. In
one-dimensional structures such defects are layers with different
characteristics. Depending upon which parameters of the defect layer
are modified one can have a variety of defect structures. Effects of
such defects on reflection/transmission properties have been studied
for both
regular\cite{Impurity_modes:1993,Liu:1997,Figotin:1998,tunable_defect}
and resonant photonic crystals.\cite{DefectMQW,CitrinDefect,
DeychPLA,DefectMQWOL,OmegaDefectPRB,Optical_Properties_MQWPC,DefectAPL}
In the particular case of Bragg MQW structures it was demonstrated,
for instance, that by using different types of defects one can
engineer structures with a wide variety of optical
spectra.\cite{DefectMQW} It is clear that defects will also
substantially affect luminescence properties of the Bragg
structures. Despite the obvious interest of this issue for
applications, it has not yet been addressed, and in this paper we
present the initial analysis of this problem for one particular type
of the defect structure.

The role of the defects on the luminescence spectrum of Bragg MQW
structures is two-fold. Firstly, the defects affect the emission of
the regular part of the structure caused by the modification of the
transmission spectrum, $t_N$, of the structure. Secondly, the defect
layers, depending on their structure, can contribute their own
luminescence to the total spectrum. It should be noted, however,
that the second contribution is expected to be small because of the
much smaller number of the defect layers compared to the total
number of the periods in the structure.

In this paper we will illustrate a possibility to modify
luminescence of Bragg MQW structures with the help of structure
manipulation by considering one particular case of a defect
structure.  We will consider a $(2N + 1)$-layer multiple quantum
well structure, in which the first and the last $N$ layers have a
fixed width $w$, while $(N+1)$-th layer at the middle does not
contain a quantum well and has a width $d$. To simplify our analysis
we will neglect the refractive index contrast between well and
barriers in the structure. This type of defect can be described as a
cavity, in which parts of the structure to the right and to the left
of the defect layer are identified as mirrors. Transmission and
reflection properties of this structure in the region of the
stop-band have been studied in Ref.~\onlinecite{DefectMQW} in the
limit of very long structures. Here we will consider more realistic
case of relatively short structures, and will analyze the
manifestations of this defect in luminescence.

The transmission properties of the structure are described by the
transfer matrix
\begin{equation}\label{eq:defect_structure_transfer}
  T = T_w^N T_d T_w^N,
\end{equation}
where $T_w$ and $T_d$ are transfer-matrices through layers with and
without quantum wells, respectively,
\begin{equation}\label{eq:transfer_layer_with_well}
  T_w =\begin{pmatrix}
   e^{i \phi_w}(1 - i S) & -i S \\
   i S & e^{-i \phi_w}(1 + i S)
 \end{pmatrix},
 \qquad
   T_b =\begin{pmatrix}
   e^{i \phi_d} & 0 \\
   0 & e^{-i \phi_d}
 \end{pmatrix},
\end{equation}
with $\phi_w = k w$ and $\phi_d = kd$.

The luminescence spectrum can be calculated using general expression
(\ref{eq:field_spectral_density}). As has been noted, the strongest
effect on the luminescence spectrum can be expected due to the
modification of the transmission by the defect layer. We consider
this effect for the case when it is most pronounced, i.e. when MQW
parts of the structure satisfy the Bragg condition,
$\phi_w(\omega_0) = \pi$. In $N$-layer MQW structures without the
defect the transmission in this case has a deep with the width $\sim
N\Gamma_0$ centered at the exciton frequency, $\omega_0$. The
resonant tunneling induced by the defect mode results in the
appearance of the resonant transparency that, in turn, leads to
resonant grow of the luminescence at the respective frequencies. In
order to provide a qualitative description of this effect it is more
convenient to work with reflection $r = -T_{21}/T_{22}$ and consider
its resonance drop caused by the defect layer. Using the
representation (\ref{eq:theta_beta_representation}) for the transfer
matrices surrounding the defect layer, one can find for not too long
structures
\begin{equation}\label{eq:T_12_defect_structure}
  T_{21} \propto \cos\phi_d +
  N\sin\phi_d (\sin\phi_w - S \cos \phi_w).
\end{equation}
Taking into account that $T_{22}$ does not have a significant
frequency dependence, we can assume that the minimum of the
reflection, $\omega_{min}$, occurs at the same frequency  as the
minimum of $T_{21}$. Neglecting the homogeneous width of the exciton
resonance compared to the width of the stop-band $\Delta_\Gamma$, we
can present an equation determining $\omega_{min}$ in the form:
\begin{equation}\label{eq:minimizing_reflection_defect}
 \frac{\omega_{min}}{\omega_{d}} = 1 - \frac 1 \pi
 \arctan\left(\frac{\omega_{min} - \omega_0}{N\Gamma_0}\right),
\end{equation}
where $\omega_d$ is defined by $\phi_d(\omega_d) = \pi$. %It should
%be noted that the extremum curve given by
%Eq.~(\ref{eq:minimizing_reflection_defect}) corresponds to maximum
%as $\omega_{min} = \omega_0$.
\begin{figure}
  \includegraphics[width=6.5in]{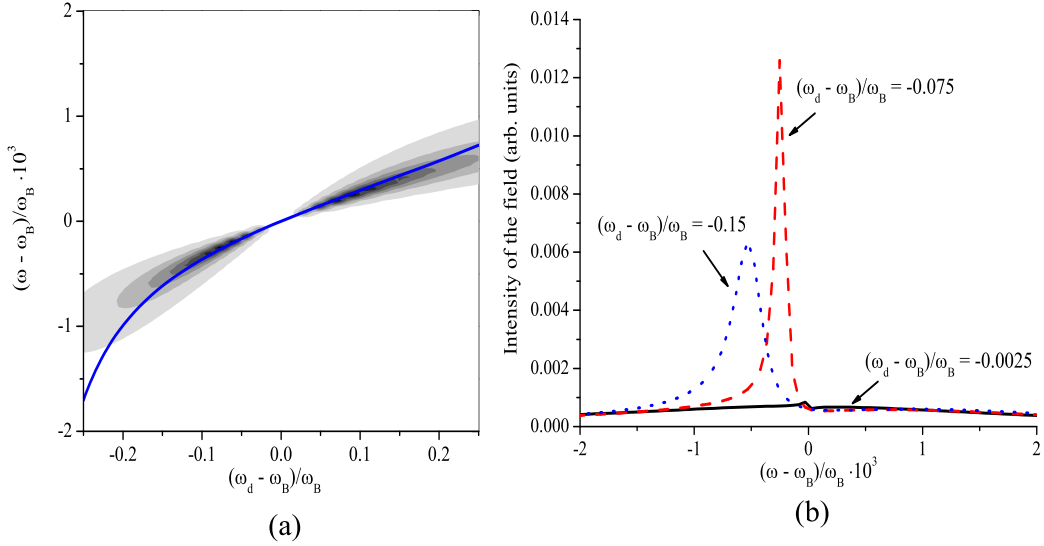}
  \caption{(color online)(a). The luminescence spectrum of the Bragg $40$ layers structure with the defect
  layer inserted at the middle of the structure. The intensity of
  the emitted field is plotted as a function of frequency (the vertical
  axis) for structures with the defect layer characterized by
  different $\omega_d$ (the horizontal axis). The solid line
  depicts the position of the maximum of the transmission as
  obtained in Eq.~(\ref{eq:minimizing_reflection_defect}). The
  parameters of the structure are the same as in
  Fig.~\ref{fig:luminescence_band_structure_fine}a. (b). Explicit
  form of the luminescence spectrum of the defect structure for
  several values of the defect resonant frequency $\omega_d$.
 }\label{fig:lum_defect_phase}
\end{figure}

As follows from this consideration, the defect layer inserted into a
MQW structure leads to the appearance of a transparency resonance
with the width $\sim \gamma$. The position of the maximum of the
emission is determined by the interplay between the maximum of the
transmission and the decay of the exciton luminescence at
frequencies away from $\omega_0$. The maximum of the field
intensity, therefore, is expected at frequencies shifted from
$\omega_{min}$ towards $\omega_0$.

In Figure~\ref{fig:lum_defect_phase} we plot the luminescence
spectrum for the structures with different width of the defect
layer. The spectrum is obtained by performing direct summation in
Eq.~(\ref{eq:field_spectral_density}) using the invariant embedding
method\cite{DEYCH:1999} for finding the partial transfer matrices.
As one can see, the numerical calculations confirm the simple
analysis provided above and demonstrate a sharp rise in the
luminescence due to the defect state.

\subsection{Absorption and luminescence spectra}\label{sec:L&A}

According to the Kirchhoff's law all emitting system in
thermodynamic equilibrium must demonstrate an universal relationship
between their absorption and emission spectra. Such a relationship
exists even in non-equilibrium but stationary situations such as
luminescence in the regime of steady state excitation. To establish
such a relationship for a particular system, however, is not always
a straightforward task. For instance, authors of
Ref.~\onlinecite{Cavity_Luminescence:1996} managed to establish such
a relationship for quantum well excitons only as an approximation.
At the same time, this relationship might be rather important
because it could allow to relate to each other various microscopic
characteristics of a system under consideration independently of
particular microscopic model used for their calculation. Here we
will use the approach to description of resonant photonic crystals
developed in the present paper as well as in
Refs.~\onlinecite{Optical_Properties_MQWPC, DispersionMQWPC} in
order to derive \emph{an exact} relation between absorption and
emission spectra of resonant photonic crystals.

A traditional definition of absorbance, $A(\omega)$, of an open
one-dimensional dielectric structure has the form
\begin{equation}\label{eq:abs_via_transm}
A(\omega)=1-T(\omega)-R(\omega)
\end{equation}
where the last two terms represent transmission and reflection
coefficients respectively. This form can be effectively used to
carry out a thermodynamical derivation of the relation between
emission and absorption specifically tailored for one-dimensional
systems. Let's assume that our one-dimensional structure is bounded
by vacuum on its left-hand side and by a homogeneous dielectric
medium with refractive index $n_b$ on its right-hand side. The
derivation is based on the statement that in equilibrium total
photon flux out of the system must be equal to zero. This total flux
includes the flux of emitted photons, $\Phi_{em}$, and the flux of
the incident and transmitted/reflected photons, $\Phi_{i}$,
$\Phi_{t}$, and $\Phi_{r}$, respectively. $\Phi_{em}$ can be
calculated as
\begin{equation}\label{eq:flux_emitted}
\Phi_{em}=I_0(\omega)d\omega\frac{dq_xdq_y}{(2\pi)^3},
\end{equation}
where $I_0$ is the equilibrium luminescence intensity, and $q_{x,y}$
represent conserving in-plane components of the photons' wave
vector. The incoming and reflected fluxes in vacuum can be presented
as
\begin{equation}\label{eq:vacuum_flux}
\begin{split}
\Phi_i+\Phi_r=\hbar\omega
cN_{ph}^0(\omega)\left[1-R(\omega)\right]\frac{dq_xdq_ydq_z}{(2\pi)^3}\\
=\hbar\omega
N_{ph}^0(\omega)\left[1-R(\omega)\right]d\omega\frac{dq_ydq_z}{(2\pi)^3}\:,
\end{split}
\end{equation}
where $N_{ph}^0(\omega)$ is the equilibrium photon occupation
number. The last contribution to the flux of photons in vacuum comes
from the photons transmitted from the medium on the right-hand side
of the structure. This flux can be written down as
\begin{equation}\label{eq:trans_flux}
\Phi_t=\frac{c}{n_b}N_{ph}^0(\omega)\tilde{T}(\omega)\frac{dq_xdq_ydq_z}{(2\pi)^3}\\
=\hbar\omega
N_{ph}^0(\omega)\tilde{T}(\omega)d\omega\frac{dq_ydq_z}{(2\pi)^3}
\end{equation}
where $\tilde{T}$ stands for the transmission coefficient of light
incident on the system from the right-hand side, and we took into
account the refractive index $n_b$ of the medium on the right.
Combining Eq.(\ref{eq:flux_emitted}) with
Eq.(\ref{eq:vacuum_flux}),(\ref{eq:trans_flux}) and with the
requirement of the zero total flux we can write down
\begin{equation}
\hbar\omega
N_{ph}^0(\omega)\left[1-R(\omega)-\tilde{T}(\omega)\right]=I_0(\omega)
\end{equation}
Taking into account Eq.(\ref{eq:abs_via_transm}) and the fact that
due to the time reversal symmetry transmission coefficients,
$\tilde{T}$ for the wave incident from right is equal to the one
describing waves incident from left we obtain a final form of the
Kirchhoff's law for one-dimensional layered structures
\begin{equation}\label{eq:Kirchhoff_law}
I_0(\omega)=\hbar\omega N_{ph}^0(\omega)A(\omega)
\end{equation}
Equilibrium photon distribution function at low temperatures $k_B T
\ll \hbar \omega$ can be approximated by $\exp{(-\hbar \omega/ k_B
T)}$.

The derivation of Eq.~(\ref{eq:Kirchhoff_law}) is based on two main
assumptions: zero photon flux, and time reversal symmetry, both of
which are valid for any kind of steady state, not necessarily
equilibrium, situations. In non-equilibrium cases, however, the
concept of photon distribution function, $N_{ph}(\omega)$, appearing
in Eq.(\ref{eq:Kirchhoff_law}), becomes ambiguous, and
generalization of this equation to these situations is not
straightforward. Nevertheless, under certain circumstances, a
relation similar to Eq.(\ref{eq:Kirchhoff_law}) can be obtained for
such non-equilibrium phenomena as luminescence under steady state
excitation. Results of numerous studies of exciton
photo-luminescence in QW (see, for instance,
Ref.\onlinecite{Ivchenko:Book, Chemla:1990}) indicate that in the
case of non-resonant photo-excitation of luminescence at the
moderately low temperatures where the main contribution to
luminescence already comes from free excitons, the following kinetic
of luminescence can be assumed. Originally excited electron-hole
pairs first relax through phonon-assisted processes to high energy
states with in-plane wave numbers $k$ corresponding to non-radiative
"dark" excitons\cite{Chemla:1990}. These states live long enough to
come in quasi-equilibrium with the crystal lattice, so that they can
be characterized by a Boltzmann distribution $f_B(E)\propto
\exp{[-(E-\mu)/k_BT]}$. The dark exciton act as a source of exciton
luminescence, and, according to calculations presented in Appendix,
the photoluminescence intensity $I(\omega)$ is a linear functional
of the exciton distribution function, $f_B(E)$. Term $\exp{(\mu/k_B
T)}$ in the Boltzmann distribution can be factored out, and the
remaining expression reproduces equilibrium emission intensity. Thus
for the intensity of the luminescence we can write $I(\omega) =
\exp{(\mu/k_B T)} I_0(\omega)$, from which it follows that
$I(\omega)$ and $A(\omega)$ are related to each other as
\begin{equation}\label{eq:nonKirchhoff_law}
I(\omega)=\hbar \omega \exp{(\mu/k_B T)} N_{ph}^0(\omega)A(\omega)
\end{equation}

Experimental verification of the relation given by
Eq.(\ref{eq:nonKirchhoff_law}) can provide useful qualitative
information about the distribution of excitons and their kinetics
either this relation is confirmed or not.
Eq.(\ref{eq:nonKirchhoff_law}) can be applied to emission and
absorption of each QW constituting the structure, and if the
distribution functions of excitons are identical in each well, to
the entire MQW structure as well.

The established relation between emission and absorption is based on
thermodynamical arguments and depends little on the details of the
structures under consideration. It appears useful to compare this
relation with the one derived on the basis of solution of Maxwell
equations for the particular model of MQW structure considered in
this paper. To this end it is more convenient to use an alternative
expression for absorption, which follows directly from the
definition of Poynting vector and energy conservation:
\begin{equation}\label{eq:absorption_def}
  A = -\frac{ \oint \mathbf{S}\cdot d\mathbf{a}}{S_0 L_xL_y}
\end{equation}
where $\mathbf{S}$ is the Poynting vectors of outgoing radiation
respectively,  $S_0$ and $L_x$, $L_y$ are the magnitude of the
Poynting vector of the incoming radiation and the transverse
dimensions of the sample respectively; the integral is taken over a
surface enclosing the entire sample. The particular convenience of
Eq.~(\ref{eq:absorption_def}) for 1D structures stems from the fact
that this expression allows for expressing the absorption in terms
of the fields at the boundaries of the structure. Assuming that the
incoming radiation impinges on the structure from the left, where
the space is filled with the medium with refractive index $n_L$, one
can obtain for the absorption coefficient
\begin{equation}\label{eq:absorption_1D}
  A = \left.\frac{1}{2ik_L\mid E_0\mid^2}\left(E \frac{dE^*}{dz} - E^*
  \frac{dE}{dz}\right)\right|_{z_L}^{z_R},
\end{equation}
where $k_L$ is the wave number of the field in the surrounding
medium to the left of the sample, and $E_0$ is the electric field
amplitude of the incoming radiation. Multiplying
Eq.~(\ref{eq:field_scalar_equation}) without sources and its
conjugate by $E^*$ and $E$, respectively, and integrating over the
entire structure one obtains
\begin{equation}\label{eq:absorption_field}
  A = \frac{4\pi\omega^2}{k_Lc^2}\sum_m \left|\int_{QW} dz\, \Phi_m(z)
  E(z)\right|^2\mathrm{Im}\,\chi_m.
\end{equation}
This expression has clear physical meaning. It shows that the
absorption of the resonant photonic crystal is the sum of
independent contributions of all quantum wells. Each contribution
has an expected form of the product of the imaginary part of the
exciton susceptibility and a term proportional to the projection of
the em field onto the exciton state in the well. In order to find a
relation between the absorption and the emission spectra we
calculate the absorption of the wave of unit intensity incident
normally from the left. Electric field inside each well, and
respective integrals in Eq.(\ref{eq:absorption_field}), can be found
using standard transfer-matrix technique (see, for instance
Ref.~\onlinecite{{JointComplex},{Optical_Properties_MQWPC}}). As a
result Eq.~(\ref{eq:absorption_field}) can be presented in the form
explicitly containing the same Green's functions as one used in this
paper, Eq.(\ref{eq:Green_functions_definition}):
\begin{equation}\label{eq:absorption_from_left}
  A =  \frac{c}{n_L\pi\omega} \sum_m \frac{\mathrm{Im}\,\chi_m}{|\chi_m|^2}
  \left|\frac{4\pi\omega^2}{c^2}\tilde{\chi}_m\right|^2
  \left|\mathcal{G}^{(s)}(m){\varphi_m}_1 +
  \mathcal{G}^{(a)}(m){\varphi_m}_2\right|^2.
\end{equation}
Comparing this expression with Eq.~(\ref{eq:field_outside_Green})
(with terms proportional to $F_1$ and $F_2$ omitted) we can relate
the emitted intensity to the absorption coefficient for a single
$m$-th well of the structure
\begin{equation}\label{eq:absorption_vs_emission}
  \mathcal{I}_-(m) = \frac{\pi n_L\omega}{c}\frac{\Xi_m|\chi_m|^2}
  {\mathrm{Im}\,\chi_m}A(m)= \frac{\pi n_L\omega}{c}\frac{\Xi_m\alpha}
  {\gamma}A_(m)
\end{equation}
where, in the second expression, we assumed that exciton
susceptibility has a Lorentzian form and is described by
Eq.~(\ref{eq:modified_susceptibility}). This assumption can be
violated under several circumstances, for instance, when several
exciton levels spectrally overlap and contribute to the
emission,\cite{EREMENTCHOUK:2005} or in the presence of the
inhomogeneous broadening of excitons. If the latter case is treated
in the effective medium approximation, as was discussed previously
in this paper and in
Refs.~\onlinecite{{OmegaDefectPRB},AndreaniKavokinIB}, it results in
an effective  susceptibility with non-Lorentzian shape.

In order to derive a global relation between emission and absorption
for the entire structure, Eq. (\ref{eq:absorption_vs_emission}) has
to be summed over all wells. If all wells are identical, i. e. the
source functions and non-radiative decay rates in all wells are the
same, the summation is trivial and the global relation between
absorption and emission coefficients has again the form of Eq.
(\ref{eq:absorption_vs_emission}). Comparing this result with
Eq.(\ref{eq:nonKirchhoff_law}) we can establish relationship between
microscopic parameters such as the strength of the exciton-light
coupling, characterized by $\alpha$, non-radiative decay rate
$\gamma$, the source function $\Xi$, which is proportional to
polarization correlation function, and photon distribution function,
$N_{ph}^{(0)}$:
\begin{equation}\label{eq:microsc_relation}
\hbar N_{ph}^{(0)}(\omega)= \frac{\pi}{c}\exp{(-\mu/k_B
T)}\frac{\Xi(\omega)\alpha}{\gamma(\omega)}
\end{equation}
Since  the photon distribution function is practically independent
of frequency on the scale of frequencies considered in this paper,
Eq.(\ref{eq:microsc_relation}) is consistent with an assumption that
both $\gamma$ and $\Xi_m$ are frequency independent quantities.
Taking into account also that $1/\omega$ term in
Eq.(\ref{eq:absorption_vs_emission}) also changes very weakly on the
same scale, we can conclude on the basis of both
Eq.(\ref{eq:Kirchhoff_law}) and Eq.(\ref{eq:absorption_vs_emission})
that emission and absorption spectra in our case are directly
proportional to each other: $\mathcal{I}_-(\omega)\propto
A(\omega)$. Numerical evaluation of the respective expressions
completely confirm this conclusion.

The assumption of all wells being the same, however, may violate in
photoluminescent experiments due to attenuation of the pumping
radiation. As a result, different wells may be characterized by
different exciton distribution functions, and consequently by
different source functions. In this case, while
Eq.(\ref{eq:absorption_vs_emission}) remains valid locally for any
particular well, its global version is not true anymore. This
results in loss of the direct proportionality between absorption and
emission spectra of our structures. This is clearly seen from
Fig.\ref{fig:pump_absorp}, which shows modification of emission
intensity caused by attenuation of pump, while absorption spectra
obviously are not affected by this circumstance.
\section{Conclusion}

In the present paper we studied spectrum of non-coherent radiation
emitted by one-dimensional resonant photonic crystal structures.
While for concreteness we focused on exciton luminescence in
multiple-quantum-well structures, the general theoretical framework
developed in this work can be applied to other structures of this
sort. The results obtained in the paper can be classified in two
groups. First, we have developed a powerful method of solving
general linear response type of problems for one-dimensional layered
structures of general type. The problem of luminescence of
one-dimensional resonant photonic structures is just one example of
such problems, in which one is looking for the radiative response of
the system caused by incoherent periodically distributed emitters.
Our approach allows expressing Greens' function of the structure in
terms of transfer matrices describing propagation of the radiation
through the system. As a result we are able to present the spectrum
of the emitted light in terms of reflection and transmission
coefficients of the structure in question. Also, with the help of a
special version of transfer-matrix description of
transmission/reflection properties of resonant photonic crystals
developed in recent Refs.~\onlinecite{DispersionMQWPC} and
\onlinecite{Optical_Properties_MQWPC} we were able to obtain a
closed analytical expression for the spectrum of luminescence of an
resonance photonic crystal structure with an arbitrary number of
identical periods. An important characteristics of these general
results is that they are obtained in terms of particular solutions
of an initial value problem for a structure with an arbitrary
spatial profile of the refractive index. The latter problem can
always be easily solved either analytically, or in most cases at
least numerically, and, therefore, emission characteristics are
expressed in our approach in terms of easily accessible quantities.

The second group of the results is concerned with application of our
general formalism to the particular case of Bragg or near-Bragg
multiple-quantum-well structures.  We analyzed the luminescence
spectrum of these structures and established its main qualitative
and quantitative characteristics. In particular, we explained the
absence of luminescence in the spectral region of polariton
stop-band, which was shown to be due to a combination of two
factors: diminishing of transmission of light through the structure
in the vicinity of the exciton frequency, and a significant spectral
separation of the latter from transparent regions because of
formation of the wide band-gap. It is interesting to note that this
result agrees with quantum-field calculations of
Ref.\onlinecite{KIRA:2006}, where similar effect was called
"subradiance". Our calculations show, however, that this effect can
be explained on purely classical ground.

We also considered modification of the spectrum when the period of
the structure becomes slightly de-tuned from the exact Bragg
conditions. Comparison of our calculations with experimental spectra
demonstrated an important role played by inhomogeneous broadening of
excitons in formation of the spectra of luminescence of the
structures under consideration. We also showed that these spectra
are influenced by a great deal of other effects such as attenuation
of the pump, or presence of cladding layers in the structure. We
found that our calculations produce good agreement with experiment
in terms of positions of the peaks of the luminescence, but not for
the relative height of the peaks. This, however, is not very
surprising, because the intensities of the peaks depend on many
various circumstances. One of the most intriguing effect, which was
not considered in the paper, but which could significantly influence
the distribution of the luminescence intensity between its peaks, is
acoustic phonon-induced scattering between collective exciton
polariton states formed in quasi-Bragg MQW structures. This
possibility is supported by the fact that spectral separation
between luminescence peaks in typical
experiments\cite{HUBNER:1999,{MINTSEV:2002}} is of the same order of
magnitude ($1 meV$) as an average energy of acoustic phonons in
$GaAs$. Consideration of this effect is out of the scope of the
current paper, but will be presented in the subsequent publications.

In order to achieve a better understanding of luminescence of the
quasi-Bragg structures, one needs additional experimental data,
which would provide information for assessing the role of different
effects. For instance, since one can, to some extent, control
inhomogeneous broadening of excitons by growth conditions,
luminescence measurements on a series on sample grown under
different conditions could clarify the role of inhomogeneous
broadening. Another possibility can be to excite luminescence by
pumping the sample from both sides, which will reduce effects due to
inhomogeneity of source function, and assess the role of this
effect. Temperature dependence of the intensity of the luminescent
maxima can be used to verify the role of acoustic phonon scattering.

An interested question studied in the last section of the paper is
concerned with relation between luminescence and absorption spectra
in multiple-quantum-well structures. First, using thermodynamical
arguments we derived a version of Kirchhoff's law specifically
adapted for one-dimensional structures under consideration. Then,
using developed formalism, we were able to establish the relation
between luminescence and absorption in terms of microscopic
characteristics of the system such as polarization correlation
function and exciton susceptibility. Comparing the two results we
established a relation between the microscopic parameters consistent
with the Kirchhoff's law. In particular, we found that if the
spectral region of interest is small compared to the characteristic
energy scale of the photon distribution function, both polarization
correlation function and the non-radiative decay rate of excitons
can be considered as frequency independent.

\acknowledgments The work by Ioffe Institute's group was supported
by the RFBR and programmes of the RAS. The Queens College group
would like to acknowledge partial support of AFOSR via grant
F49620-02-1-0305, as well as support by PCS-CUNY grants. Work at the
Northwestern University was partially supported by NSF grant No. DMR
0093949.

\section*{Appendix}\label{sec:Appendix}
In this Appendix we will provide a sketch of quantum-mechanical
calculations of luminescence from a single quantum well. The
objective of this exercise is to provide a microscopic justification
for the quasi-classical approach employed in the paper, and
demonstrate relations between phenomenological source function
$\Sigma(\rho,t)$ and microscopical characteristics of the system.
This relation can be derived only by using the second quantization
of the exciton and photon states and a relevant microscopic kinetic
equation. Here we will demonstrate the second-quantization approach
for a single QW-structure with a QW layer sandwiched between
semi-infinite barriers. For simplicity, the dielectric contrast $n_w
- n_b$ is set to zero.

The 2D ${\bm k}$-space is naturally divided into two regions,
radiative and nonradiative, respectively, with $k < k_0 \equiv
(\omega_0/c) n_b$ and $k> k_0$. In the process of relaxation of
electron-hole pairs non-radiative excitons with $k> k_0$ are being
populated first\cite{Chemla:1990}, and we assume that, for these
excitons the criterion $\bar{k} l \gg 1$ for validity of the
Boltzmann kinetic equation is fulfilled. Here ${\bar{k}}^2$ is the
average value of $k^2$ describing the spread of exciton population
in the ${\bm k}$-space, $l$ is the 2D exciton mean free-path length,
$l = (\hbar \bar{k}/M) \tau_p$, $\tau_p$ is the exciton momentum
scattering time, and $M$ is the exciton in-plane effective mass. For
3D photons we introduce a quantization box of the volume $V=SL$ with
the macroscopic interface area $S$ and the macroscopic length $L$
along the QW growth direction.

\begin{figure}
  \includegraphics[width=\linewidth]{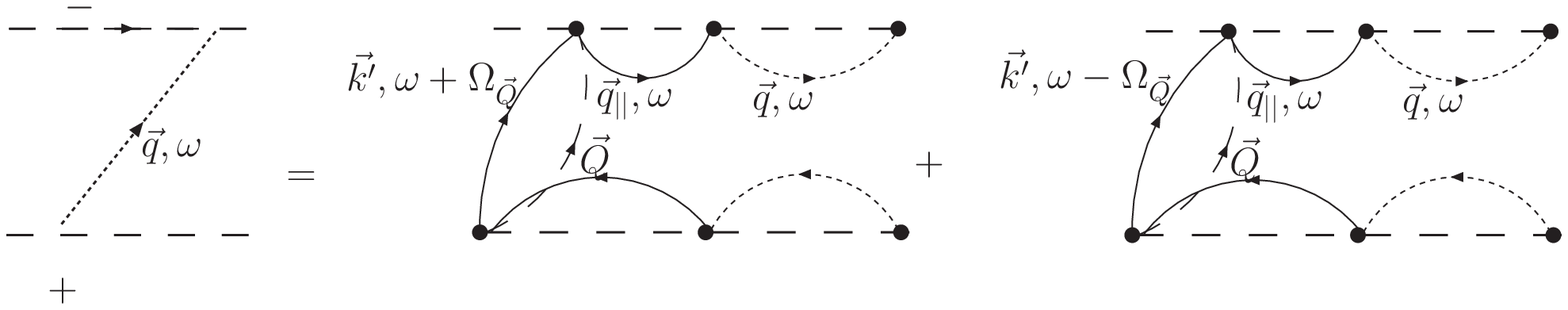}
  \caption{.}\label{fig:e}
\end{figure}

In order to calculate the spectral intensity of light emitted by
excitons in the QW we apply the Keldysh diagram
technique\cite{LifshitsPitaevsky,{Keldysh:1964}}, in a way similar
to one used for the description of exciton or exciton-polariton
photoluminescence in bulk crystals, see for instance, Ref.~
\onlinecite{Ivchenko:1977,{Ivchenko:1989}}. In this method the
intensity $I_{\bm q}$ can be written as
\begin{equation} \label{Iq}
I_{\bm q} = \hbar \omega_{\bm q} w_{\bm q}\:,\:w_{\bm q}= -
\lim_{\gamma \to + 0} \left( \frac{\gamma}{\pi} \int D^{- +}_{{\bm
q}, \omega}~ d \omega\right)\:.
\end{equation}
Here $w_{\bm q}$ is the emission rate of a photon with the 3D photon
wave vector ${\bm q}$, $D^{- +}_{{\bm q}, \omega}$ is the photon
Green function presented in the left-hand side of the diagram
equation shown in Fig.~\ref{fig:e} by an short-dashed line
connecting the lower and higher horizontal parts of the Keldysh
contour labeled "+" and "-", respectively; $\gamma$ is a positive
photon damping rate which is introduced for the formal reasons in
order to stabilize the photon distribution in the steady-state
regime [$(2 \gamma)^{-1}$ is the photon lifetime in the quantization
box]. In the end $\gamma$ is set to $+ 0$ since in a single QW
structure with the microscopic length $L$ the actual photon damping
rate is negligible. Equation (\ref{Iq}) is derived taking into
account that, in the steady-state regime, one has
\begin{equation} \label{D-+}
D^{- +}_{{\bm q}, \omega} = - 2 \pi N_{\bm q} \left( \frac{1}{\pi}
\frac{\gamma}{(\omega - \omega_{\bm q})^2 + \gamma^2}\right)\:,
\end{equation}
where $\omega_{\bm q} = cq/n_b$, $N_{\bm q} = w_{\bm q}/(2 \gamma)$
is the steady-state photon distribution function. Here the
temperature is assumed to be very small as compared to the exciton
excitation energy, which allows one to neglect the equilibrium
photon-state population. Note that the close-to-normal energy flux
in the frequency range $d \omega$ and within the area $d^2
q_{\parallel}$ in the plane $(q_x, q_y)$ is related with $I_{\bm q}$
by
\[
dI(\omega) = \frac{ d^2 q_{\parallel} d \omega}{(2 \pi)^3}
\frac{Ln_b}{c} I_{\bm q}\:.
\]

The Green function $D^{- +}_{{\bm q}, \omega}$ is found from the
diagram equation presented in Fig.~\ref{fig:e}. The diagrams on the
right-hand side of the equation describe the acoustic-phonon
assisted scattering of an exciton from the non-radiant state ${\bm
k}'$ to the radiative state ${\bm q}_{\parallel}$, where ${\bm
q}_{\parallel}$ is the in-plane component of ${\bm q}$; ${\bm Q}$
and $\omega_{\bm Q}$ are the phonon wave vector and frequency. In
Fig.~\ref{fig:e} the short-dashed lines mean ${\rm i} D^{s's}_{{\bm
q} \omega}$, where the superscripts $s', s = \pm$ show the position
of the ends of the photon Green functions,
\[
D^{--}_{{\bm q} \omega} = - \left( D^{++}_{{\bm q} \omega} \right)^*
= \frac{1}{\omega - \omega_{\bm q} + {\rm i} \gamma}\:.
\]
The solid lines represent exciton Green functions ${\rm i}
G^{s's}_{{\bm k} \omega}$, given by
\[
G^{- +}_{{\bm k}', \omega \pm \Omega_{\bm Q}} = - 2 \pi f_{{\bm k}'}
\left( \frac{1}{\pi} \frac{ \Gamma_{{\bm k}'} }{ (\omega \pm
\Omega_{\bm Q} - \omega^{({\rm exc})}_{{\bm k}'})^2 + \Gamma_{{\bm
k}'}^2} \right)\:,
\]
\[
G^{--}_{{\bm k} \omega} = - \left( G^{++}_{{\bm k} \omega} \right)^*
= \frac{1}{\omega - \omega^{({\rm exc})}_{\bm k} + {\rm i}
(\Gamma_{\bm k} + \Gamma_{0{\bm k}})}\:,
\]
$\omega^{({\rm exc})}_{\bm k}, \Gamma_{\bm k}$ and $\Gamma_{0{\bm
k}}$ are the exciton excitation energy, non-radiative and radiative
damping rates, $f_{{\bm k}'}$ is the exciton distribution function
assumed to be small, $f_{{\bm k}'} \ll 1$. Each vortex represents
the matrix element of exciton-phonon ($V_{\bm Q}$) or exciton-photon
($M_{{\bm q}_{\parallel}}$) interaction multiplied by $\pm {\rm i}/
\hbar$, plus for the lower part and minus for the upper part of the
Keldysh contour. Since we neglect the damping of acoustic phonons
the upward and downward phonon (long-dashed) lines can simply be
replaced by the factors $m_{\bm Q}$ and $m_{\bm Q} + 1$, where
$m_{\bm Q}$ is the phonon occupation number. For a process described
by any vortex the conservation of the in-plane wave vectors should
be satisfied. In particular, ${\bm k}' = {\bm k} \pm {\bm
Q}_{\parallel}$ for the phonon emission and absorption,
respectively. As a result, we obtain
\[
{\rm i} D^{- +}_{{\bm q}, \omega} = \frac{1}{\hbar^4} \sum_{{\bm
k}', {\bm Q}} \vert D^{--}_{{\bm q} \omega} \vert^2 \vert
G^{--}_{{\bm q}_{\parallel} \omega} \vert^2 \vert M_{{\bm
q}_{\parallel}} \vert^2 \vert V_{\bm Q} \vert^2
\]
\[
\times {\rm i} [G^{- +}_{{\bm k}', \omega - \Omega_{\bm Q}} m_{\bm
Q}\ \delta_{{\bm k}', {\bm k} - {\bm Q}_{\parallel}} + G^{- +}_{{\bm
k}', \omega + \Omega_{\bm Q}} (m_{\bm Q} + 1)\ \delta_{{\bm k}',
{\bm k} + {\bm Q}_{\parallel}}]\:.
\]
where summation over $k\prime$ is carried only over non-radiative
states  because the population of the radiative states is assumed to
be negligible. It follows then that the photon generation rate
$w_{\bm q}$ in Eq.~(\ref{Iq}) can be presented in the form
\begin{equation} \label{finalwq}
w_{\bm q} = \frac{W^{({\rm phot})}(\omega_{\bm q}, {\bm
q}_{\parallel})\ W^{({\rm phon})}(\omega_{\bm q}, {\bm
q}_{\parallel}) }{2 (\Gamma_{{\bm q}_{\parallel}} + \Gamma_{0{\bm
q}_{\parallel}})}\:,
\end{equation}
\[
W^{({\rm phot})}(\omega, {\bm k}) = \frac{2 \pi}{\hbar^2} \vert
M_{{\bm k}} \vert^2\ \frac{1}{\pi} \frac{ \Gamma_{ {\bm k} } +
\Gamma_{ 0 {\bm k} } }{ (\omega - \omega^{({\rm exc})}_{{\bm k}})^2
+ ( \Gamma_{ {\bm k} } + \Gamma_{ 0 {\bm k} } )^2}\:,
\]
\[
W^{({\rm phon})}(\omega, {\bm k}) = \frac{2 \pi}{\hbar^2} \sum_{{\bm
k}', {\bm Q}} \vert V_{\bm Q} \vert^2 f_{{\bm k}'} [m_{\bm Q}\
\delta(\omega - \Omega_{\bm Q} - \omega^{({\rm exc})}_{{\bm k}'})\
\delta_{{\bm k}', {\bm k} - {\bm Q}_{\parallel}}
\]
\[
 +\ (m_{\bm Q} + 1)\ \delta(\omega +
\Omega_{\bm Q} - \omega^{({\rm exc})}_{{\bm k}'})\  \delta_{{\bm
k}', {\bm k} + {\bm Q}_{\parallel}}]\:.
\]
Here, taking into account the condition $\bar{k} l \gg 1$ we
replaced in the last equation the phonon's resonance Lorentzians by
Dirac delta-functions.

In the Bragg and quasi-Bragg structures, the radiative excitonic
states $| n, {\bm k} \rangle$ in different wells $n$ with the same
wave vector ${\bm k}$ are strongly coupled by the electromagnetic
field. In contrast, excitons with ${\bm k}$ satisfying the
conditions $k > k_0$, $\sqrt{k^2 - k_0^2} d \gg 1$  are decoupled
and can be considered as excitations isolated in particular wells.
The photon emission outgoing from a given QW can be described by
Eq.~(\ref{finalwq}) as well or, equivalently. On the other hand, the
propagation of this outgoing light wave through the whole MQW
structure accompanied by its reabsorption and escape into the vacuum
or substrate, can be described classically.

Instead of the above microscopical consideration, in the main part
of the paper the photoluminescence spectra are calculated with the
help of a quasi-classical Langevin-like approach by introducing the
random source $\Sigma_m({\bm \rho}, \omega)$ into expression for
exciton polarization, see
Eq.~(\ref{eq:exciton_polarization_singleQW}). Now, the explicit
expression for the correlator $\Xi_m(\omega, {\bm k})$ of the random
sources in Eq.~(\ref{eq:spectral_density}) can be readily found from
Eq.~(\ref{finalwq}). In particular, one has
\begin{equation} \label{xiomega0}
\Xi_m(\omega, 0) = \frac{2 \pi q}{\hbar \varepsilon_b \Gamma_0^2}
|M_0|^2 W^{(\rm phon)}(\omega, 0) \left[ \int \Phi_m(z)dz \right]^2
\:,
\end{equation}
where $\Gamma_0$ is the radiative damping rate of an exciton with
${\bm k}=0$ excited in a single QW structure.

The applicability of the Langevin method can be justified by
considering Heisenberg equations of motion for exciton annihilation
($\hat{c}_{\bm k}$) and creation ($\hat{c}^{\dagger}_{\bm k}$)
operators. For the exciton-phonon Hamiltonian
\[
\sum_{{\bm k} {\bm k}' {\bm Q}} \hat{c}^{\dagger}_{\bm k}
\hat{c}_{{\bm k}'} \left( \hat{d}_{\bm Q} V_{\bm Q} \delta_{{\bm
k}', {\bm k} - {\bm Q}_{\parallel}} + \hat{d}^{\dagger}_{\bm Q}
V_{\bm Q}^* \delta_{{\bm k}', {\bm k} + {\bm Q}_{\parallel}}
\right)\:,
\]
where $\hat{d}_{\bm Q}, \hat{d}^{\dagger}_{\bm Q}$ are the
acoustic-phonon annihilation and creation operators, one obtains
\begin{equation}
\dot{\hat{c}}_{\bm k}(t) = - \omega^{({\rm exc})}_{\bm k}
\hat{c}_{\bm k}(t) - \frac{\rm i}{\hbar} M_{\bm k}^* \hat{b}_{\bm
k}(t) + \hat{\xi}_{\bm k}(t)
\end{equation}
with $\hat{b}_{\bm k}$ being the photon annihilation operator. Here
term
\[
\hat{\xi}_{\bm k}(t) = - \frac{\rm i}{\hbar} \sum_{{\bm k}' {\bm Q}}
\left( V_{\bm Q} \hat{d}_{\bm Q} \delta_{{\bm k}', {\bm k} - {\bm
Q}_{\parallel}} + V_{\bm Q}^* \hat{d}^{\dagger}_{\bm Q} \delta_{{\bm
k}', {\bm k} + {\bm Q}_{\parallel}} \right) \hat{c}_{{\bm k}'}
\]
playes the role of the Langevin source operator, and summation again
is carried over only values of $k\prime$ corresponding to
non-radiative states. Fourier transform of the correlator of this
operator
\[
{\rm i} \int d \tau\ \langle \hat{\xi}^{\dagger}_{\bm k}(t)
\hat{\xi}_{\bm k}(t + \tau) \rangle {\rm e}^{{\rm i} \omega \tau}
\]
is equal to $W^{({\rm phon})}(\omega, {\bm k})$, which agrees with
Eq.~(\ref{xiomega0}) if we take into account the relation between
the 2D exciton envelope function and exciton-induced polarization
$P_{\rm exc}$.

%\bibliographystyle{apsrev}
%\bibliography{luminescence_paper}
%\bibliography{optical_properties}

\end{document}